\documentclass[11pt]{article}

\usepackage{verbatim}
\usepackage{amsmath}
\usepackage{amsfonts}
\usepackage{amssymb}
\usepackage{graphicx}
\usepackage{bm}
\usepackage{color}
\usepackage{mathrsfs}
\usepackage{epsf}
\usepackage[hypertex]{hyperref}

\begin{document}

\title{Global properties of Stochastic Loewner evolution driven
by L\'evy processes}
\author{P. Oikonomou, I. Rushkin, I. A. Gruzberg, L. P. Kadanoff
\\ \\
The James Franck Institute, The University of Chicago\\
5640 S. Ellis Avenue, Chicago, Il 60637 USA }

\date{January 14, 2008}

\maketitle

\begin{abstract}

Standard Schramm-Loewner evolution (SLE) is driven by a
continuous Brownian motion which then produces a trace, a
continuous fractal curve connecting the singular points of the
motion. If jumps are added to the driving function, the trace
branches. In a recent publication [1] we introduced a
generalized SLE driven by a superposition of a Brownian motion
and a fractal set of jumps (technically a stable L\'evy
process). We then discussed the small-scale properties of the
resulting L\'evy-SLE growth process. Here we discuss the same
model, but focus on the global scaling behavior which ensues as
time goes to infinity. This limiting behavior is independent of
the Brownian forcing and depends upon only a single parameter,
$\alpha$, which defines the shape of the stable L\'evy
distribution. We learn about this behavior by studying a
Fokker-Planck equation which gives the probability distribution
for endpoints of the trace as a function of time. As in the
short-time case previously studied, we observe that the
properties of this growth process change qualitatively and
singularly at $\alpha =1$. We show both analytically and
numerically that the growth continues indefinitely in the
vertical direction for $\alpha > 1$, goes as $\log t$ for
$\alpha = 1$, and saturates for $\alpha< 1$. The probability
density has two different scales corresponding to directions
along and perpendicular to the boundary. In the former case,
the characteristic scale is $X(t) \sim t^{1/\alpha}$. In the
latter case the scale is $Y(t) \sim A + B t^{1-1/\alpha}$ for
$\alpha \neq 1$, and $Y(t) \sim \ln t$ for $\alpha = 1$.
Scaling functions for the probability density are given for
various limiting cases.

\end{abstract}

\newpage

\tableofcontents

\newpage

\section{Introduction}
\label{sec:intro}

The study of random conformally-invariant clusters that appear
at critical points in two-dimensional statistical mechanics
models has been made rigorous with the invention of the
so-called Schramm-Loewner evolution (SLE) \cite{Schramm
original}. SLE refers to a continuous family of evolving
conformal maps that specify the shape of a part of a critical
cluster boundary. By now SLE has been justly recognized as a
major breakthrough, and there are several review papers and one
monograph devoted to this beautiful subject, see Refs.
\cite{lawler,werner-lectures,Schramm-review2006,
bauer-bernard-review,cardy-review,gruzberg-review,
gruzberg-kadanoff-review,kager-nienhuis-review}

SLE describes a curve, called trace, growing with time from a
boundary in a two-dimensional domain which is usually chosen to
be the upper half plane. SLE is based on the Loewner equation
in which the shape of the growing curve is determined by a
function of time $\xi(t)$ which in SLE is taken to be a scaled
Brownian motion. Such a choice of the driving function produces
continuous stochastic, fractal and conformally invariant
curves, --- the kind that appears as the scaling limit of
various interfaces in many two-dimensional critical lattice
models and growth processes of statistical physics. Well-known
examples include boundaries of the Fortuin-Kastelyn clusters in
the critical $q$-state Potts model, loops in the $O(n)$ model,
self-avoiding and loop-erased random walks.

In Ref. \cite{previous paper} we generalized SLE to a broader
class for which $\xi(t)$ is a Markov process with
discontinuities. More specifically, we have studied the Loewner
evolution driven by a linear combination of a scaled Brownian
motion and a symmetric stable L\'evy process. The growing curve
then exhibits branching. This generalized process might be
useful to describe many tree-like growth processes, such as
branching polymers and various branching growth processes which
evolve in time.

Such generalized SLEs driven by L\'evy processes (L\'evy-SLE
for short) have also been of interest to mathematics community.
Our results \cite{previous paper} on various phase transitions
in L\'evy-SLE have been put on rigorous basis in Ref.
\cite{Guan-Winkel}, and further properties have been studied in
Refs. \cite{Guan,Chen-Rohde}. The interest of mathematicians in
these L\'evy-SLE processes is partially motivated by the
suggestion \cite{Chen-Rohde,Beliaev-Smirnov-SLE-spectrum} that
they may produce fractal objects with large values of
multifractal exponents for harmonic measure. Harmonic measure
can be thought of as the charge distribution on the boundary of
a conducting cluster. On fractal boundaries such a distribution
is a multifractal, and in the case of critical clusters (whose
boundaries are SLE curves) the full spectrum of multifractal
exponents has been obtained analytically, see Refs.
\cite{Beliaev-Smirnov-SLE-spectrum,HM-Duplantier-PRL,
HM-Duplantier-long,HM-Eldad-PRL,HM-Ilia-JPhysA} for various
derivations and discussion.

While our previous paper \cite{previous paper} focused on local
properties of L\'evy-SLE, here we study the global behavior of
the growth in the upper half plain. The present paper is
structured as follows. In Section \ref{sec:model} we define our
model and briefly state our previous results on phase
transitions in the local behavior of the model. We also present
our new results on the global behavior of L\'evy-SLE. In
Section \ref{sec:FPE} we derive the Fokker-Planck equation
governing the evolution of the probability distribution for the
tip of the L\'evy-SLE. The equation is our main tool for
analysis of the long time global behavior of the growth. We
give a qualitative description of the growth and explain the
approximations that go into the solution of the Fokker-Planck
equation in Section \ref{sec:approximations}. Actual solution
of the Fokker-Planck equation and comparison with results from
numerically calculated trajectories is given in Section
\ref{sec:solution}. We conclude in Section
\ref{sec:conclusions}. Some technical details are presented in
Appendices.

\section{The model and the results, old and new}
\label{sec:model}

\begin{figure}
\centering
\includegraphics*[width=\textwidth]{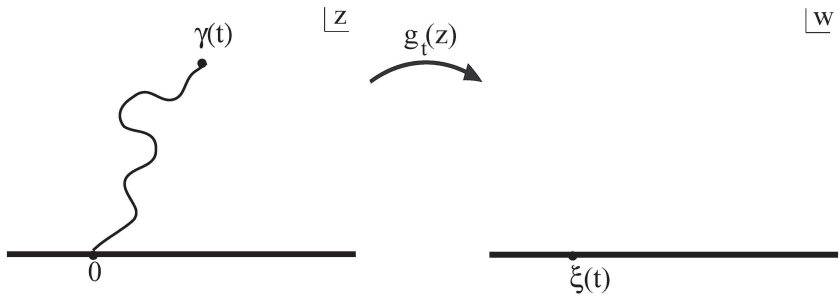}
\caption{The Loewner evolution shown for the
case when the growing hull is a smooth curve. The complement of the
segment of the curve (up to its tip $\gamma(t)$) in the ``physical''
$z$ plane is mapped to the entire upper half of the ``mathematical'' $w$
plane by the function $g_t(z)$.} \label{fig:LE}
\end{figure}

Loewner evolution is a family of conformal maps that appears as
the solution of the Loewner differential equation (see, for
example, Ref. \cite{lawler} for details)
\begin{align}
\partial_t g_t(z) &= \frac{2}{g_t(z)-\xi(t)}, & g_0(z)=z.
\label{g}
\end{align}
valid at any point $z$ in the upper half plane until (and if)
this point becomes singular at some (possibly infinite) time
$\tau_z$: $\xi(\tau_z) = g_{\tau_z}(z)$. The set of all
singularities is called the hull and the point at which the
hull grows is called the tip. The tip $\gamma(t)$ is defined
via its image $\xi(t) = g_t(\gamma(t))$. More formally,
\begin{align}
\gamma(t)=\lim_{w \to \xi(t)}g_t^{-1}(w),
\label{tip-g}
\end{align}
where the limit is taken in the upper half plane. The trace is
the path left behind by the tip (the existence of the trace in
the setting of this paper has been shown in Ref. \cite{Guan}).
The shape of the growing trace (and the hull) is completely
determined by the driving function $\xi(t)$. At any time the
function $g_t(z)$ conformally maps the exterior of the growing
hull to the upper half plane, see Fig. \ref{fig:LE}. We refer
to the $z$ plane where the growth occurs as ``the physical
plane'', and to the $w$ plane as ``the mathematical plane''.

Naturally, if $\xi(t)$ is a stochastic process the shape of the
growing trace is also stochastic. The growth process is then a
stochastic (Schramm-) Loewner evolution (SLE). The standard SLE
has a driving function $\xi(t)=\sqrt\kappa B(t)$, where $B(t)$
is a normalized Brownian motion and $\kappa>0$ is the diffusion
constant. Many important properties of this process have been
established in Ref. \cite{Rohde-Schramm}.

In Ref. \cite{previous paper} we have generalized SLE to
\begin{align}
\xi(t) = \sqrt\kappa B(t)+c^{1/\alpha}L_\alpha(t),
\label{SLEUL}
\end{align}
where $L_\alpha(t)$ is a normalized symmetric $\alpha$-stable
L\'evy process \cite{apple,metz,ta,Sato}, and $c>0$ is the
``diffusion constant'' associated with it. The process
$L_\alpha(t)$ is composed of a succession of jumps of all
sizes. Unlike a Brownian motion, $L_{\alpha}(t)$ is
discontinuous on all time-scales. Therefore, the addition of a
L\'evy processes to the driving force of SLE introduces
branching to the trace.

The probability distribution function of
$c^{1/\alpha}L_\alpha(t)$ is given by the Fourier transform
\begin{align}
P(x,t) = \int_{-\infty}^\infty \frac{dk}{2\pi}
e^{-ikx}e^{-ct|k|^\alpha}. \label{levypdf}
\end{align}
As it is known in the theory of stable distributions
\cite{Sato}, only for $0< \alpha \leqslant2$ this Fourier
transform gives a non-negative probability density. For
$0<\alpha<2$ the function $P(x,t)$ decays at large distances as
a power law:
\begin{align}
x\to\infty:\quad P(x,t)\sim\frac{ct}{|x|^{1+\alpha}},
\end{align}
so that the process scales as
\begin{align} \label{scalescale} \langle
|L_\alpha(t)|^\delta\rangle\propto t^{\delta/\alpha}
\end{align}
for any $\delta < \alpha$. For $\delta \geqslant \alpha$ this
average is infinite. For $\alpha=2$ the process $L_2(t)$ is the
standard Brownian motion $B(t)$ and $P(x,t)$ is Gaussian.

We studied the short-distance properties of the L\'evy-SLE
process in Ref. \cite{previous paper}. At short times and
distances the process is dominated by the Brownian motion and
the deterministic drift term (see Eq. (\ref{E2})), whereas at
long times it is dominated by L\'evy flights. The crossover
between short and long time behavior happens at the time
\begin{align}
t_0 \sim \Big(\frac{1}{c^2}\Big)^{1/(2 - \alpha)}. \label{CHAR-TIME}
\end{align}
This also defines a spatial crossover at length scales $l_0
\propto \sqrt{t_0}$. For scales smaller than $l_0$ the trace
behaves like standard SLE, while for scales much larger than
$l_0$ it spreads in the $x$ direction forming tree-like
structures.

\begin{figure}
\vskip -10mm
\centering
\vskip -10mm
\includegraphics*[scale=.4]{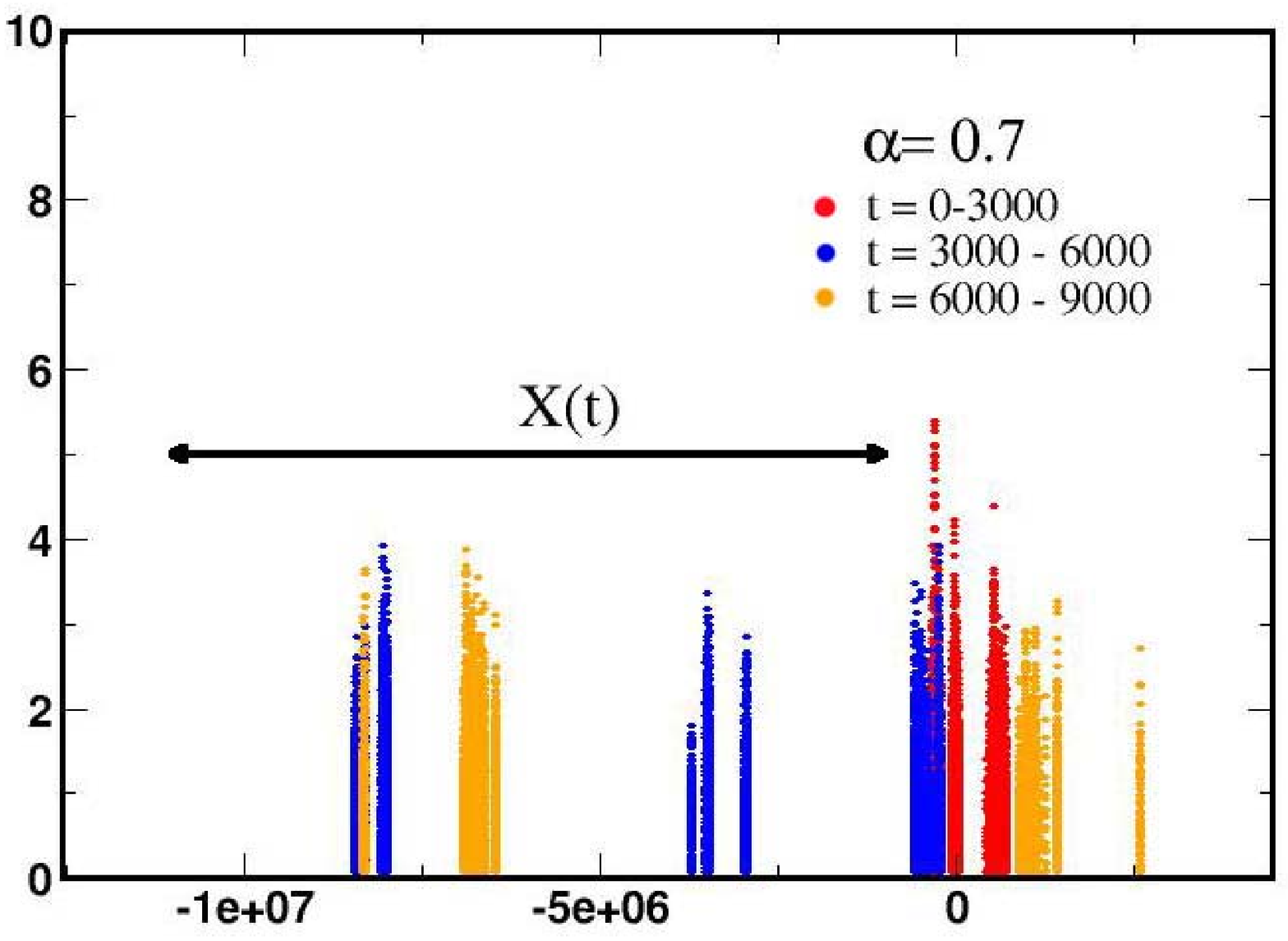}
\vskip -5mm
\includegraphics*[scale=.4]{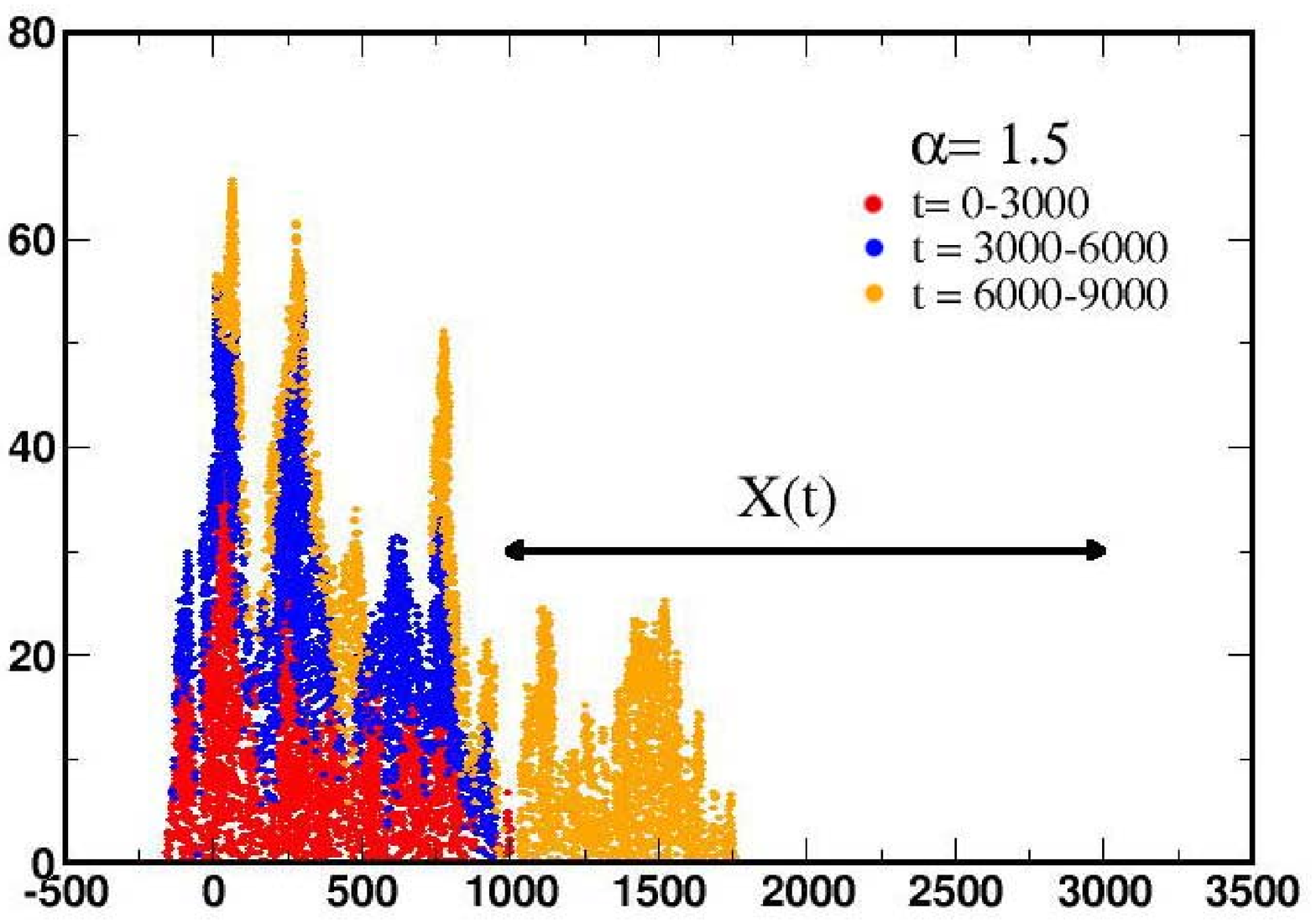}
\vskip -3mm \caption{Examples of traces produced by L\'evy-SLE at
long times, up to $t=9000$. For the first, second and last thirds of
the time interval, the traces are correspondingly colored red, blue,
and yellow. Top panel: $\alpha = 0.7$. The trace looks like many
isolated trees whose height saturates at long times. Bottom panel:
$\alpha = 1.5$. Now the tip of the growing trace keeps landing on
the previously grown ``bushes'' so that the trace extends
indefinitely in the vertical direction as time increases. Notice the
difference in scales for the $y$ axis between the two panels, as
well as the much larger spread in the $x$ direction in the top
panel. The trace was produced using stable L\'evy forcing and
$c=10$, time step $\tau= 10^{-3}$. The trace was calculated only at
times when the forcing makes a large jump $d\xi>\sqrt{200\tau}$.}
\label{fig:trees}
\end{figure}

In our previous paper \cite{previous paper}, using both
analytic and numerical considerations, we determined the
probability that a point on the $x$ axis is swallowed by the
trace. The trace shows a qualitative change in its
small-distance, small-time behavior as $\kappa$ and $\alpha$
each pass though critical values, respectively at four and one.
The transition at $\kappa=4$ is quite analogous to the known
transition of standard SLE \cite{Rohde-Schramm}. For the new
transition at $\alpha=1$, the trace forms isolated trees when
$\alpha<1$ or a dense forest when $\alpha>1$.

The latter phase transition at $\alpha = 1$ was recently
studied rigorously in Ref. \cite{Guan-Winkel} which expanded
the implications of the phase transition to the whole plane at
the limit $t\rightarrow \infty$. For $\kappa>4$ a point in the
upper half plane is swallowed almost surely for $\alpha>1$,
while it is swallowed with probability smaller than one for
$\alpha<1$. For $\kappa<4$ and $0<\alpha<2$ the swallowed
points on the plane form a set of measure zero.

The large-scale implications of the $\alpha=1$ transition can
be seen in Figure \ref{fig:trees} which shows the shape of the
trace at long times.  For $\alpha<1$ the stochastic evolution
produces isolated tree-like structures which are limited in
height. For $\alpha>1$ the evolution produces an ``underbrush''
in which structures pile on one another and thereby continue to
increase their height.

In the rest of the paper we establish the following. The growth
at long times is characterized by two very different length
scales $X(t)$ and $Y(t)$ (with $X(t) \gg Y(t)$) which can be
thought of as the typical size of the growing hull in the $x$
and $y$ directions. More specifically, we find that
\begin{align}
 X(t) &\sim t^{1/\alpha}, \quad 0 < \alpha < 2, \\
 Y(t) &\sim
  \begin{cases}
   A + B t^{1-1/\alpha}, & \alpha \neq 1, \\
   \ln t, & \alpha = 1.
  \end{cases}
\end{align}
(The constants $A$ and $B$ depend upon $\alpha$.) These scales
enter the scaling form of the joint probability distribution
$\rho(x,y,t)$ for the real and imaginary parts of the tip
$\gamma(t)$ of the L\'evy-SLE, for which we give explicit
results in various limiting cases in Section
\ref{sec:solution}, where we also compare analytical results
with extensive numerical simulations.

\section{Derivation of the Fokker-Planck equation}
\label{sec:FPE}

We are interested in characterizing the probability
distribution for the point $\gamma(t)$ at the tip of the trace
in the ensemble provided by different realizations of the SLE
stochastic process. Eq. (\ref{tip-g}) implies then that we
should study the inverse map $g_t^{-1}$. However, this is
rather difficult, since the map $g_t^{-1}$ satisfies a partial
differential equation instead of an ODE. There is a way out
which is rather well known and has been successfully used
before
\cite{Chen-Rohde,Beliaev-Smirnov-SLE-spectrum,Rohde-Schramm}.
It happens that one needs to consider the backward time
evolution:
\begin{align}
\partial_t f_t(w) &= - \frac{2}{f_t(w)-\xi(t)}, & f_0(w) = w.
\label{LE-f}
\end{align}

The relation of the original Loewner evolution (\ref{g}) and
the backward one (\ref{LE-f}) in the stochastic setting is as
follows. If $\xi(t)$ is a symmetric (in time) process with
independent identically distributed increments, which is the
case for a L\'evy process, then it is easy to show that for any
fixed time $t$ the solution $f_t(w)$ of the backward equation
(\ref{LE-f}) has the same distribution as $g_t^{-1}(w - \xi(t))
+ \xi(t)$, see Refs. \cite{Chen-Rohde, Rohde-Schramm}. Using
the symbol $\stackrel{d}{=}$ for equality of distributions for
random variables, we can write
\begin{align}
f_t(w) \stackrel{d}{=} g_t^{-1}(w - \xi(t)) + \xi(t).
\label{back-forward-equiv}
\end{align}

It is useful to introduce a shifted conformal map
\begin{align}
h_t(z) = g_t(z) - \xi(t),
\end{align}
for which the Loewner equation acquires the Langevin-like form:
\begin{align}
\partial_t h_t(z) &= \frac{2}{h_t(z)}-\partial_t\xi(t), & h_0(z)=z,
\label{E2}
\end{align}
assuming that $\xi$ vanishes at $t=0$. The first term is a
deterministic drift and the second --- a random noise. The tip
$\gamma(t)$ is now mapped to zero, and this can be taken as the
definition of the tip. More formally,
\begin{align}
\gamma(t)=\lim_{w \to 0}h_t^{-1}(w)
\label{tip}
\end{align}
where the limit is taken in the upper half plane.

In terms of the shifted map the equality of distributions
(\ref{back-forward-equiv}) can be written as
\begin{align}
f_t(w) - \xi(t) \stackrel{d}{=} h_t^{-1}(w ).
\label{back-forward-equiv-1}
\end{align}
The left hand side $z_t \equiv f_t(w) - \xi(t)$ of this
equation satisfies the Langevin-like equation
\begin{align}
\partial_t z_t &= - \frac{2}{z_t}-\partial_t\xi(t), & z_0 = w,
\label{Langevin-phys}
\end{align}
and in particular, if we set $w=0$ in this equation, the
resulting stochastic dynamics should be the same as that of the
tip of the trace $\gamma(t)$.

\begin{figure}[t]
\centering
\includegraphics*[width=0.8\textwidth]
{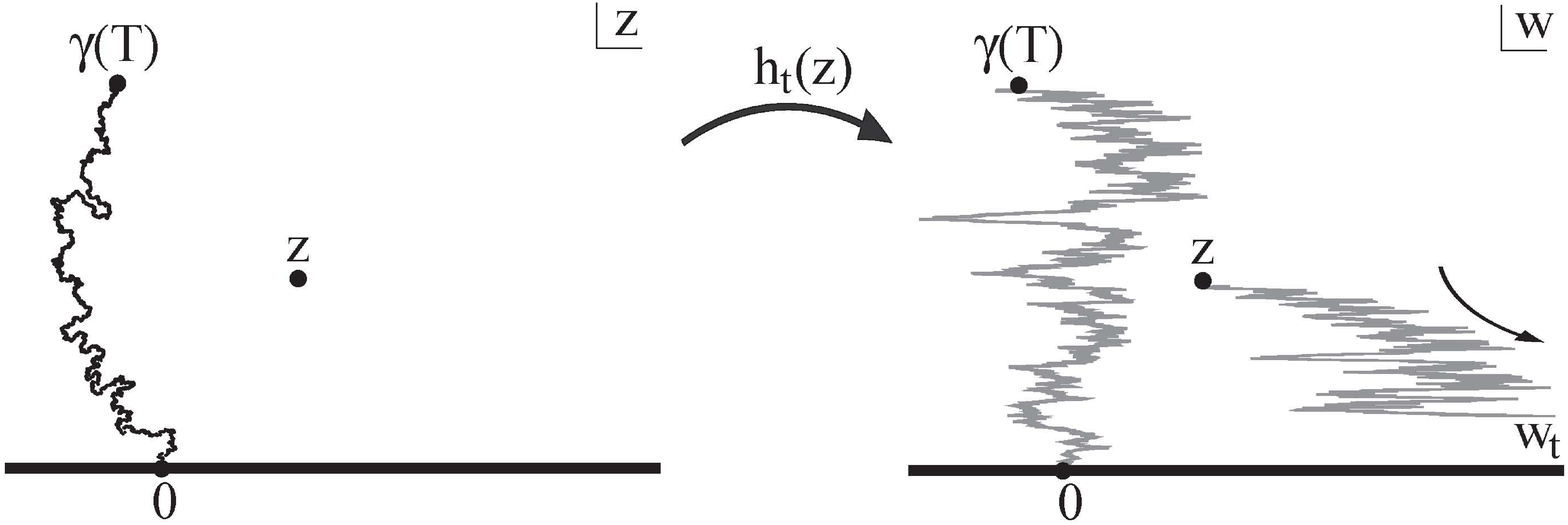}\\
\includegraphics*[width=0.8\textwidth]
{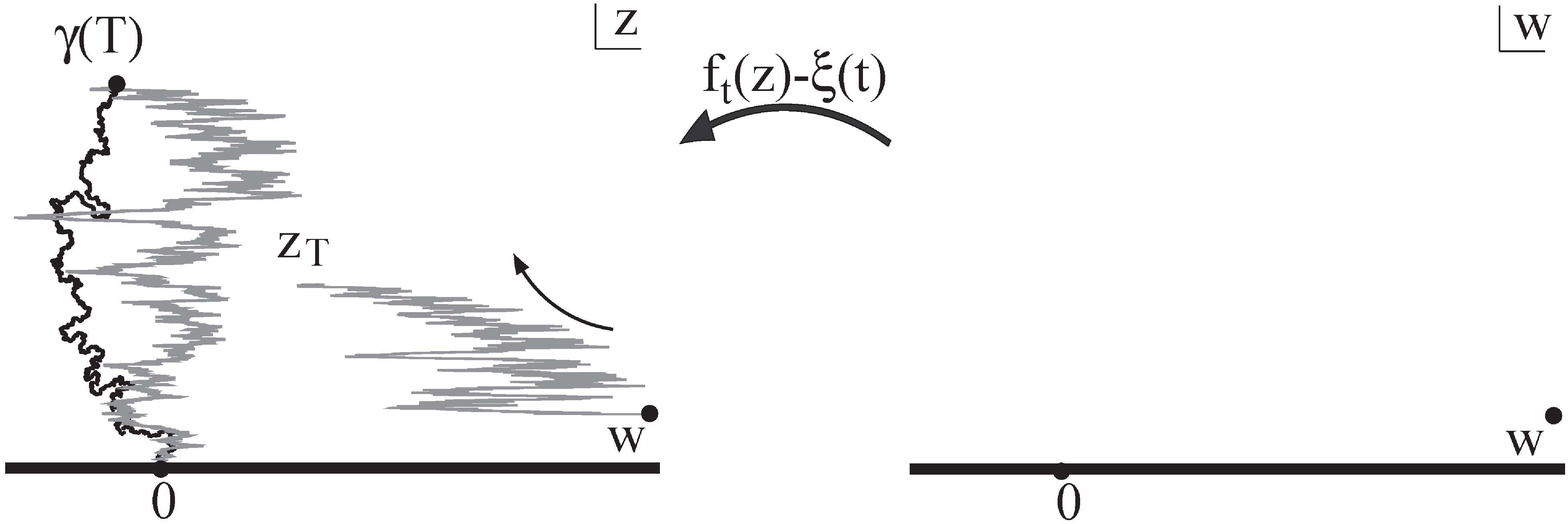} \caption{Top panel: two
trajectories in the forward flow. Bottom panel: the corresponding
trajectories in the backward flow. Curved arrows indicate the flow
of time. In both cases the flow trajectories are shown in grey,
while the black line represents an SLE trace. The trajectories on
the bottom panel have been produced with the same noise realization
as used for the forward evolution, reversed in time as described in
the main text. The grey trajectories in the top and bottom panel are
thus identical.} \label{fig:flows}
\end{figure}
Before we convert the Langevin-like equation
(\ref{Langevin-phys}) to our main analytical tool, the
corresponding Fokker-Planck equation, let us review again the
correspondence between the forward and backward flows and
illustrate it with figures. Equation (\ref{E2}) describes a
flow in which  $w_t = h_t(z)$ follows a trajectory of a
particle in the $w$ plane, $z$ being its initial position.
Separating the real and imaginary parts of $w_t = u_t + i v_t$,
we get a system of coupled equations
\begin{align}
\partial_t u_t &= \frac{2u_t}{u_t^2 + v_t^2} - \partial_t \xi(t),
& u_0 &= x, \nonumber \\
\partial_t v_t &= -\frac{2v_t}{u_t^2 + v_t^2}, & v_0 &= y,
\label{math langevin}
\end{align}
describing such a trajectory. As in many modern versions of
dynamics, all initial conditions and hence all trajectories are
considered at the same time, forming an ensemble. Two such
trajectories are presented on the top panel in Fig.
\ref{fig:flows}. For a generic initial point $z$ the trajectory
$w_t$ goes to infinity in the horizontal $u$ direction, while
the vertical coordinate $v_t$ monotonously decreases. However,
if the initial point happens to be a point $\gamma(T)$ on the
SLE trace, the forward trajectory hits the origin in the
mathematical plane exactly at time $T$.

Conversely, we can fix a point $w$ in the mathematical plane
and follow the motion of its image $z_t$ under the map $f_t(w)
- \xi(t)$ in the physical plane, with the initial condition
$z_0 = w$. In components $z_t = x_t + i y_t$, the trajectories
of this backward flow satisfy the system of equations
\begin{align}
\partial_t x_t &= -\frac{2x_t}{x_t^2 + y_t^2} - \partial_t \xi(t),
& x_0 &= u, \nonumber \\
\partial_t y_t &= \frac{2y_t}{x_t^2 + y_t^2}, & y_0 &= v.
\label{phys langevin}
\end{align}
The two trajectories shown on the bottom panel in Fig.
\ref{fig:flows} precisely retrace the trajectories of the
forward flow shown on the top panel. This has been achieved by
driving the backward evolution (\ref{phys langevin}) by the
time reversed noise $\xi(T-t) - \xi(T) \stackrel{d}{=} \xi(t)$
as compared to the forward evolution. In this case the final
point $z_T$ of the trajectory that started at the origin
coincides with the tip of the trace $\gamma(T)$ at that time,
but the rest of the trajectory does not follow the SLE trace.
If we drive the backward flow by an independent copy of
$\xi(t)$, then even the final point $z_T$ will be different
from $\gamma(T)$, but in the statistical ensemble $z_T$ and
$\gamma(T)$ will have the same distribution.

Now we can introduce the probability distribution function of
the process $z_t=x_t + i y_t$ in the physical plane defined by:
\begin{align}
\rho(x,y,t) &=\langle\delta(x_t - x) \delta(y_t - y) \rangle.
\label{definition of phys pdf}
\end{align}
From Eqs. (\ref{SLEUL}, \ref{phys langevin}) it follows
immediately that $\rho(x,y,t)$ satisfies the following
(generalized) Fokker-Planck equation:
\begin{align}
\partial_t\rho(x,y,t) = \Bigl[ \frac{\kappa}{2} \partial_x^2 +
c|\partial_x|^\alpha + \partial_x \frac{2x}{x^2+y^2} - \partial_y
\frac{2y}{x^2+y^2}\Bigr] \rho(x,y,t). \label{FPE}
\end{align}
Here $|\partial_x|^\alpha$ (sometimes also written as
$(-\Delta)^{\alpha/2}$) is the Riesz fractional derivative,
which is a singular integral operator whose action is easiest
to describe in the Fourier space: if $\tilde f(k)$ is the
Fourier transform of a function $f(x)$, then the Fourier
transform of $|\partial_x|^\alpha f(x)$ is $|k|^\alpha\tilde
f(k)$.

As we have discussed, at long times the growth is dominated by
the stable process in the driving function (\ref{SLEUL}), and
we can set $\kappa = 0$. So our main analytical tool is the
following Fokker-Planck equation:
\begin{align}
\partial_t\rho(x,y,t) = \Bigl[c|\partial_x|^\alpha + \partial_x
\frac{2x}{x^2+y^2} - \partial_y \frac{2y}{x^2+y^2}\Bigr]
\rho(x,y,t). \label{Levy-FPE}
\end{align}

Let us discuss the boundary and initial conditions for this
equation. The initial condition for the Fokker-Planck equation
(\ref{Levy-FPE}) depends on the initial conditions $x_0 = u$,
$y_0 = v$ in the stochastic equations (\ref{phys langevin}).
For the distribution of the SLE tip $\gamma(t)$ the appropriate
initial conditions are $x_0 = 0$, $y_0 = \epsilon$, where
$\epsilon$ is an infinitesimal positive number. For the exact
Fokker-Planck equation (\ref{FPE}) this translates into the
initial condition
\begin{align}
\rho(x,y, 0) = \delta(x) \delta(y - \epsilon).
\label{initial-exact}
\end{align}
However, for the approximate equation (\ref{Levy-FPE}) the
situation is more subtle. The crossover time $t_0 = O(1)$. For
$t < t_0$ the drift in the $x$ direction (towards $x=0$)
dominates over the L\'evy term. For $t > t_0$ the opposite is
true. A simple picture is then that before $t_0$ the initial
$\delta$ function is advected by the drift velocity $2/y$ in
the $y$ direction. By the time $t_0$ it becomes
\begin{align}
\rho_0(x,y) \equiv \rho(x,y, t_0) &= \delta(x)
\delta(y - y_0), & y_0= 2 t_0^{1/2} \sim c^{-1/(2-\alpha)}.
\label{initial}
\end{align}
This is the initial value that we shall assume for our problem.
In the following sections we will mostly use the notation
$\rho_0(x,y)$, using the explicit expression when necessary.

Let us comment that if we tried to be more careful and included
the effects of the Brownian forcing before the crossover time
$t_0$, then the distribution at time $t_0$ would not only be
advected to $y_0$ but would also broaden to a Gaussian with
variance $\kappa t_0$. This refinement would not change any
arguments in the later sections, since all we need there is
that the Fourier transform in $x$ of the initial distribution
is broader than $e^{-ct|k^\alpha|}$ for long times, see the
discussion preceding Eq. (\ref{approx0}). This is a good
approximation for both the initial distribution (\ref{initial})
or its Gaussian variant for sufficiently long times, and
becomes better and better as time increases.

As for the boundary conditions at $y=0$, we have no need to be
very explicit about them, since $\rho(x,y,0)$ vanishes for $y <
y_0$, and our equations of motion (\ref{phys langevin})
represent a situation in which $y_t$ continually increases as
$t$ increases, so that $\rho(x,y,t)$ will also vanish for $y <
y_0$ at all times $t>0$.

\section{Qualitative description, distance scales}
\label{sec:approximations}

In this section we analyze in qualitative terms the long-time
limit of the evolution of the tip $\gamma(t)$, by looking at
the consequences of equations (\ref{phys langevin}). According
to the discussion in the previous section, $\mathrm{Re}\,
\gamma(t)$ and $\mathrm{Im}\, \gamma(t)$ have the same joint
distribution as $x_t$ and $y_t$.

\begin{figure}
\centering
\includegraphics[scale=.4]{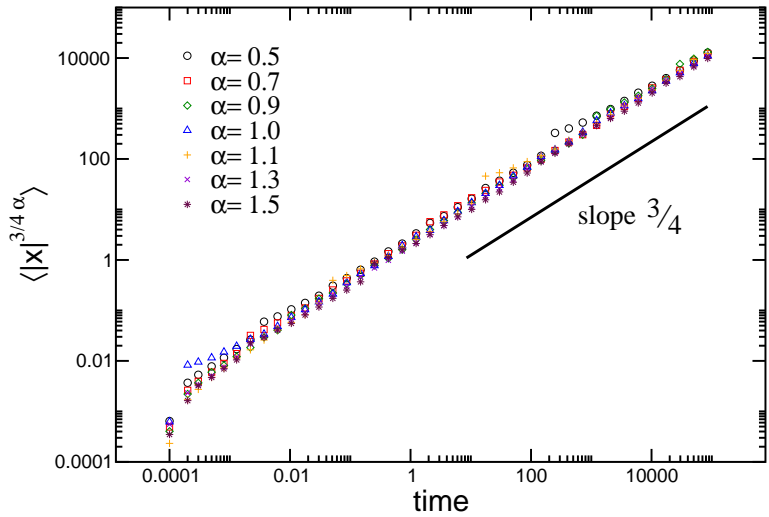}
\caption{Growth of the L\'evy-SLE parallel to the
boundary. Here we plot $\langle |x|^{3\alpha/4}\rangle$  for various
values of $\alpha$ (Brownian motion is set to zero, $\kappa= 0$).
The average follows the predicted behavior $t^{3/4}$. Data collected
by averaging realizations of equation (\ref{Langevin-phys}) for
L\'evy distributed forcing $c=1$, time step $\tau= 10^{-4}$; $10000$
runs for $\alpha= 0.5, 1.0, 1.3$, $3000$ runs for $\alpha= 0.7, 0.9,
1.1, 1.5$. The black line is a guide to the eye with the desired
slope. The irregular points observed in some of the curves are due to
large jumps in the forcing of individuals runs. Such behavior is
expected due to the power law distribution of the jumps in L\'evy
processes.} \label{fig:AvgX}
\end{figure}

For small times, up to the crossover time $t_0$, the drift term
in the Langevin equation dominates over the L\'evy noise.
Therefore, both $x_t$ and $y_t$, and $\xi(t)$ as well, grow as
$\sqrt t$. For larger times, $t \gg t_0$, there are two
different characteristic length scales, $X(t)$ and $Y(t)$. In
this regime the forcing $\xi(t)$ is dominated by the L\'evy
process $L_\alpha(t)$. The probability for a total motion
$X(t)$ over a time $t$ for this process is described by Eq.
(\ref{levypdf}). Typically the motion is dominated by a single
long jump, and the jump has an order of magnitude
\begin{align}
X(t) \sim (ct)^{1/\alpha}.
\end{align}
(This can be understood as rescaled fractional moments $\langle
|L_\alpha(t)|^\delta \rangle^{1/\delta}$, see Eq.
(\ref{scalescale})) Since the typical jumps of $\xi(t)$ become
arbitrarily large at long times, $x_t$ also becomes large, and
therefore, the drift term in the first equation of (\ref{phys
langevin}) becomes negligible. In this limit, $x_t$ behaves
like the driving force, and we find
\begin{align}
t\to\infty: |x_t| \sim X(t)\sim (ct)^{1/\alpha}.
\label{X}
\end{align}
The Loewner evolution with Levy flights produces, in general, a
forest of (sparse or dense) branching trees, growing form the
real axis. The above relation then tells us how the forest
spreads along the real axis with time. This distance is marked
out on the plots of trees shown in Figure {\ref{fig:trees}}.

Numerical implementation of the Langevin equation
(\ref{Langevin-phys}), details of which are presented in
Appendix \ref{app:numerics}, confirms these qualitative
arguments. Figure \ref{fig:AvgX} compares the estimate of Eq.
(\ref{X}) with numerical calculations of the trace via
simulations of Eq. (\ref{Langevin-phys}). The agreement is
excellent.

Next we turn to a typical distance $Y(t)$ in the $y$
coordinate. Figure \ref{fig:trees} clearly shows that this
characteristic distance is much smaller than $X(t)$. We
understand this as follows. If $x_t$ were zero, the second
equation in (\ref{phys langevin}) would give $y_t \sim
t^{1/2}$. Clearly, any non-zero $x_t$ only slows down the
growth of $y_t$. We then conclude that $y_t$, and therefore the
height of the trees produced by the SLE process cannot grow
with time faster than $t^{1/2}$. Since $\alpha < 2$, it means
that $\mathrm{Im}\,\gamma(t)$ always grows slower than
$\mathrm{Re}\,\gamma(t)$ and they become widely separated at
long times. Our major result is that  the growing trees spread
faster horizontally than they grow vertically. Hence, we have
\begin{align}
Y(t) \ll X(t). \label{YllX}
\end{align}

An estimate of the scaling of $Y(t)$ can be obtained from the
second equation in (\ref{phys langevin}) where we replace $x_t$
by the L\'evy process and average over it using the probability
distribution (\ref{levypdf}). This gives a typical behavior of
$y_t$:
\begin{align}
\partial_t y_t \approx \int_{-\infty}^\infty \!\! dx
\frac{2 y_t}{y_t^2 + x^2} P(x,t) &=
\int_{-\infty}^\infty \!\! dk \, e^{- y_t |k| - c |k|^\alpha t}.
\label{hand_waving}
\end{align}
To estimate the $k$ integral we can drop the term $y_t |k|$ in
the exponent, since this quantity is of order $Y(t)/X(t) \ll
1$. Thus we get
\begin{align}
\partial_t y_t \approx \dfrac{2\Gamma
\big(1+\frac{1}{\alpha}\big)}{c^{1/\alpha}} t^{-1/\alpha}.
\end{align}
The time integration then gives a result that the length scale
for the $y$ direction is
\begin{align}
Y(t) = y_0 + \frac{2}{c^{1/\alpha}} \frac{\Gamma(1 +
\tfrac{1}{\alpha})}{1 - \tfrac{1}{\alpha}}
t^{1-\frac{1}{\alpha}}.
\label{hand_waving2}
\end{align}
Here $y_0$ is formally the constant of integration, but it
should really be thought of as an adjustable constant inserted
to make up for any errors we might have made in doing the
integrals. In particular, it takes care of any effects from the
early-time region, where we surely do not have the calculation
under control.

\begin{figure}
\centering
\includegraphics[scale=.4]{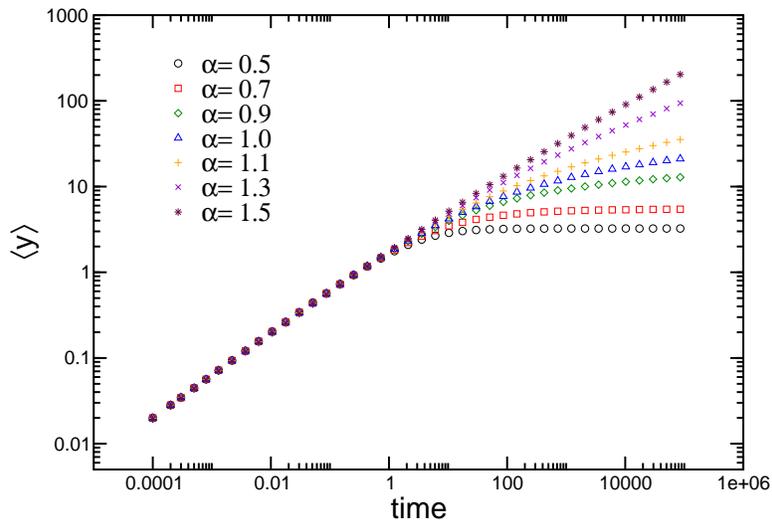}
\caption{Growth of the L\'evy-SLE perpendicular to
the boundary. We plot $\langle y\rangle$ for various values of
$\alpha$. Same details as in previous figure. Initially, the trace
grows as $\sqrt{t}$ for all values of $\alpha$. This behavior
changes around the characteristic time $t_0\sim 1$. The height of
the trace saturates for $\alpha<1$, while it grows indefinitely for
$\alpha>1$. This change of behavior demonstrates the global
implications of the phase transition at $\alpha=1$.}
\label{fig:AvgY}
\end{figure}

The phase transition at $\alpha=1$ is manifested by a
qualitative difference between $\alpha > 1$ and $\alpha < 1$.
From (\ref{hand_waving2}) we can see that while for $\alpha
\geqslant 1$ the average height of the trees grows to infinity
as $t^{1 - 1/\alpha}$, while for $\alpha <1$ it saturates at a
finite value $y_\infty$.  Figure \ref{fig:AvgY} provides an
illustration of the phase transition at $\alpha = 1$ separating
different behaviors. More detailed comparison between our
analytical predictions and numerical simulations is provided in
the next Section.

\section{Solving the FPE}
\label{sec:solution}

In order to quantify these predictions we need to return our
attention to the Fokker-Plank equation (\ref{Levy-FPE}). If we
perform the Fourier transform in $x$ and integrate in time we
can find a compact form of this equation, which reads
\begin{align}
\widetilde\rho(k,y,t) &= e^{-c|k|^\alpha t} \widetilde\rho_0(k,y) - \partial_y \int dk'
\int_0^t dt' e^{-y|k'|-c|k|^\alpha(t-t')}
\widetilde\rho(k - k',y,t') \nonumber\\
& \quad + k \int dk'\int_0^tdt' \,{\mathrm{sgn}}(k') e^{-y|k'|-c|k|^\alpha(t-t')}
\widetilde\rho(k - k',y,t'), \label{FP-integrated}
\end{align}
where $\widetilde\rho_0(k,y)$ is the Fourier transform of the
initial distribution (\ref{initial}). At long times
$\rho(x,y,t)$ is spread over the scale $X(t)$ as a function of
$x$. Its Fourier transform $\widetilde\rho(k,y,t)$, as a
function of $k$, is significantly non-zero on the scale
$X(t)^{-1}$. At the same time, due to the exponential factors
$e^{-y|k'|}$, the relevant values of $k'$ in the integrals in
Eq. (\ref{FP-integrated}) are of the order $y^{-1} \gtrsim
Y(t)^{-1}$. The scale $Y(t)^{-1}$ is much larger than the range
$X(t)^{-1}$ where $\rho$ is non-zero, hence, when integrating
over $k'$ we can use the approximation
\begin{align}
\widetilde\rho(k - k',y,t) \approx \delta(k - k') \int \!\! dk''
\widetilde\rho(k'',y,t) = 2\pi \rho(0,y,t) \delta(k - k').
\label{rho-delta}
\end{align}
The Fokker-Planck equation then reads:
\begin{align}
\widetilde\rho(k,y,t) &=  e^{- c|k|^\alpha t} \widetilde\rho_0(k,y) - 2\pi
\int_0^t \!\! dt' e^{- y|k| - c|k|^\alpha (t-t')}
\partial_y \rho(0,y,t') \nonumber\\
& \quad + 4\pi |k| \int_0^t \!\! dt' e^{- y|k| - c|k|^\alpha
(t-t')} \rho(0,y,t'). \label{approx-FP-k-y-t-space}
\end{align}
This is the main approximation that we will use in order to
study the behavior of the L\'evy-SLE process at large times.

Notice here that the distribution function $\rho$, for every
$x$ and $t$, depends only on the initial condition $\rho_0$ and
the history of the distribution at $x=0$ for earlier times
$t'<t$. Therefore, in order to study the probability density
function described by the Fokker-Planck equation, we first need
to calculate the behavior of this distribution for small $x$,
that is $\rho(0,y,t)$. Then, by substituting in Eq.
(\ref{approx-FP-k-y-t-space}), we can in principle estimate the
full distribution. However, in this paper we are mostly
interested in the way this process grows in the $y$ direction.
Hence, we will first find $\rho(0,y,t)$ which characterizes the
growth near $x=0$, and then obtain the distribution
\begin{align}
p(y,t) \equiv \int_{-\infty}^\infty \!\! dx \, \rho(x, y, t) =
\widetilde\rho(0, y, t) \label{p-y-t-definition}
\end{align}
of $y$'s integrated over all $x$ by setting $k=0$ in
(\ref{approx-FP-k-y-t-space}):
\begin{align}
p(y,t) &= p_0(y) - 2\pi \int_{0}^t \!\! dt' \, \partial_y
\rho(0, y, t'). \label{p-y-t-equation-1}
\end{align}
This equation immediately leads to the average $\langle y
\rangle$, which is understood as the average over all $x$:
\begin{align}
\langle y \rangle &= y_0 + 2\pi \int_{0}^t \!\! dt'
\int_0^\infty \!\!dy \, \rho(0,y,t'). \label{y-average}
\end{align}
Therefore, the distribution and its mean in Eqs.
(\ref{p-y-t-equation-1}) and (\ref{y-average}) depend only on
the behavior at $x=0$ at times $t'<t$. This is a direct
implication of Eq. (\ref{approx-FP-k-y-t-space}) and our main
approximation (\ref{rho-delta}).

Let us emphasize again that our approximation works in the long
time limit. We will assume that we can use approximate
expressions in time integrals for all $t > t_0$. Thus, we will
treat all time integrals $\int_0^t$ as $\int_{t_0}^t + \text{
correction}$. The corrections come from short times, and we
cannot extract them from our analysis. They all will be hidden
in the terms dependent on the lower limit $t_0$ of the time
integrals. In several cases the lower cut-off at $t_0$ is
necessary to avoid spurious divergencies.

Let us now consider $\rho(0,y,t)$. A closed equation for this
quantity results from integrating Eq.
(\ref{approx-FP-k-y-t-space}) over $k$. To do this we observe
that in the first term (the initial value at $t = 0$) for the
relevant values of $k$ the function $\widetilde\rho_0(k,y)$ is
much broader in $k$ than $e^{- c|k|^\alpha t}$ at long times.
Hence, in the integral over $k$ we can replace
$\widetilde\rho_0(k,y)$ by its value at $k=0$. Then it follows
that
\begin{align}
\rho(0,y,t) &= \frac{\widetilde\rho_0(0,y)}{2\pi X(t)} - \int_0^t \!\! dt'
\, \frac{\partial_y \rho(0, y, t')}{X(t - t',y)} - 2 \int_0^t \!\! dt'
\rho(0, y, t') \partial_y \frac{1}{X(t - t',y)}, \label{approx0}
\end{align}
where the scale $X(t,y)$ is defined as
\begin{align}
\frac{1}{X(t, y)} &= \int_{-\infty}^\infty \!\! dk \,
e^{-c|k|^\alpha t - y|k|}, \label{}
\end{align}
and
\begin{align}
X(t) = X(t, y=0) =
\dfrac{c^{1/\alpha}}{2\Gamma\big(1+\frac{1}{\alpha}\big)}
t^{1/\alpha}. \label{X-t}
\end{align}

Equation (\ref{approx0}) is easily solved after performing the
Laplace transformation in time $t$. For the transform
\begin{align}
\rho(0,y,\lambda) &
= \int_0^\infty \!\! dt \, e^{-\lambda t} \rho(0,y,t)
\end{align}
we obtain an ordinary differential equation
\begin{align}
\partial_y \rho(0, y, \lambda) + \frac{1 + 2 \partial_y K(\lambda,
y)}{K(\lambda, y)} \rho(0, y, \lambda) = \frac{K(\lambda)}{2\pi
K(\lambda, y)} \widetilde\rho_0(0, y), \label{diff-rho-0-y-lambda}
\end{align}
where
\begin{align}
K(\lambda, y) = \int_0^\infty \!\! dt \, \frac{e^{-\lambda
t}}{X(t, y)} = \int_{-\infty}^\infty \!\! dk \, \frac{e^{-y
|k|}}{\lambda + c|k|^\alpha}, \label{K-lambda-y}
\end{align}
and $K(\lambda)=K(\lambda,0)$. Using the initial condition
$\widetilde\rho_0(0, y)= \delta(y - y_0)$, the straightforward
solution of Eq. (\ref{diff-rho-0-y-lambda}) is
\begin{align}
\rho(0, y, \lambda) &= \frac{K(\lambda)}{2\pi} \frac{K(\lambda,
y_0)}{K^2(\lambda, y)} \exp \Big( - \int_{y_0}^y \!\!
\frac{dy'}{K(\lambda, y')}  \Big).
\label{solution-0-y-lambda-simplified}
\end{align}
The inverse Laplace transform of this solution gives
$\rho(0,y,t)$.

Notice that (\ref{solution-0-y-lambda-simplified}) is valid
only for $y > y_0$. Since our approximations only work at long
times, we expect our solution to give good results for $y \gg
y_0$. The approximations will usually result in the necessity
to introduce a fitting parameter (called ``correction'' in the
discussion after Eq. (\ref{y-average})) in the time evolution
of averages for the process. Moreover, there is an upper
cut-off that stems from the Langevin equation and the fact that
$y$ cannot grow faster than $t^{1/2}$ (see previous section).
Since we used this fact while making the approximations that
lead to Eq. ({\ref{approx-FP-k-y-t-space}), the range of
validity of our solution is $y_0 \ll y \ll t^{1/2}$.

In the following we will analyze the properties of the
distributions $\rho(0,y,t)$ and $p(y,t)$ in three separate
cases $\alpha>1$, $\alpha=1$ and $\alpha<1$. For each case we
will repeat the following steps: first we calculate
$\rho(0,y,t)$ from Eq. (\ref{solution-0-y-lambda-simplified}),
then, by substituting this solution into Eq.
(\ref{p-y-t-equation-1}), we will calculate the average height
$\langle y \rangle$ and the distribution $p(y,t)$. In these
calculations we need approximate expressions for the function
$K(\lambda,y)$. These expressions are derived in Appendix
\ref{app:XK-asymptotics}.

\subsection{Results for $\alpha > 1$}

In this case we can use the approximation (\ref{K-approx-a>1})
from Appendix \ref{app:XK-asymptotics} for $K(\lambda)$ and
$K(\lambda,y)$. Eq. (\ref{solution-0-y-lambda-simplified}) then
gives
\begin{align}
\rho(0, y, \lambda) &\approx \frac{1}{2\pi} \exp \Big( -
\frac{1}{A} \lambda^{1 - 1/\alpha} y  \Big), & A =
\frac{2\pi}{\alpha c^{1/\alpha} \sin\frac{\pi}{\alpha}}.
\label{solution-alpha>1}
\end{align}
To calculate the time dependence of the distribution we take
the inverse Laplace transform:
\begin{align}
\rho(0,y,t) &\approx \frac{1}{2\pi t}
\int_{a-i\infty}^{a+\infty} \frac{d\lambda}{2\pi i} e^{\lambda
t - \lambda^{1 - 1/\alpha} y/A}. \label{contour-integral}
\end{align}
As usual, the integration contour in the last equation goes
along a vertical line Re$\,\lambda = a$, where $a$ should be
greater than the real part of any singularity of the integrand.
Changing the integration variable to $\lambda t$ we obtain that
answer which, apart from the overall prefactor $1/t$, has
acquired the form of a scaling function:
\begin{align}
\rho(0,y,t) &\approx \frac{1}{2\pi t} F(\hat y), \qquad \hat y
\equiv \frac{y}{Y(t)}, \qquad Y(t) = \frac{2}{c^{1/\alpha}}
\frac{\pi}{\alpha \sin \tfrac{\pi}{\alpha}} t^{1 -
\frac{1}{\alpha}}, \label{Y-t} \\
F(\hat y) &=  \int \frac{d\lambda}{2\pi i} e^{\lambda -
\lambda^{1 - 1/\alpha} \hat y}. \label{F-y-hat}
\end{align}
Since the scaling function $F(\hat y)$ depends only on the
combination $y t^{-1 + 1/\alpha}$, its derivatives with respect
to $y$ and $t$ are related:
\begin{align}
\partial_y F(\hat y) &= - \frac{\alpha}{\alpha - 1} \frac{t}{y}
\partial_t F(\hat y).
\label{F-y-hat-derivatives-relation}
\end{align}

The integrand in Eq. (\ref{F-y-hat}) contains a branch cut
which we choose to run along the negative real axis. The
integration contour can be deformed to go from $-\infty$ to $0$
along the lower side of the cut, and then from 0 to $-\infty$
along the upper side. This leads to the final answer for the
scaling function $F(\hat y)$:
\begin{align}
F(\hat y) &= \frac{1}{\pi}
\int_0^\infty \!\! d\lambda \, e^{-\lambda - |\cos
\frac{\pi}{\alpha}| \lambda^{1 - 1/\alpha} \hat y} \sin
\Big(\sin \frac{\pi}{\alpha} \lambda^{1 - 1/\alpha} \hat y
\Big). \label{F-y-hat-integral}
\end{align}

The overall prefactor $1/t$ in $\rho(0,y,t)$ can be understood as
follows. The distribution $\rho(x, y, t)$ at long times spreads in
the $x$ direction up to scale $X(t)$, and in the $y$ direction up to
scale $Y(t)$. The total area ``covered'' by the distribution scales
with time as $X(t) Y(t) \propto t$. Therefore, at the particular
value $x=0$ the density $\rho(0, y, t)$ decays with time as $1/t$.
However, if we are looking at the distribution of the $y$ coordinate
for $x=0$, and its moments $\langle y^n \rangle$, we should divide
$\rho(0,y,t)$ by the normalization
\begin{align}
\int_0^\infty \!\! dy \, \rho(0, y, t) = \frac{Y(t)}{2\pi t
\Gamma\big(1 - \tfrac{1}{\alpha}\big)}.
\label{normalization-alpha>1}
\end{align}
The normalized distribution is then
\begin{align}
\rho_n(0,y,t) &\approx \frac{\Gamma\big(1 -
\tfrac{1}{\alpha}\big)}{Y(t)} F(\hat y).
\label{rho-0-y-t-alpha>1-normalized}
\end{align}

\begin{figure}
\centering
\includegraphics[scale=.4]{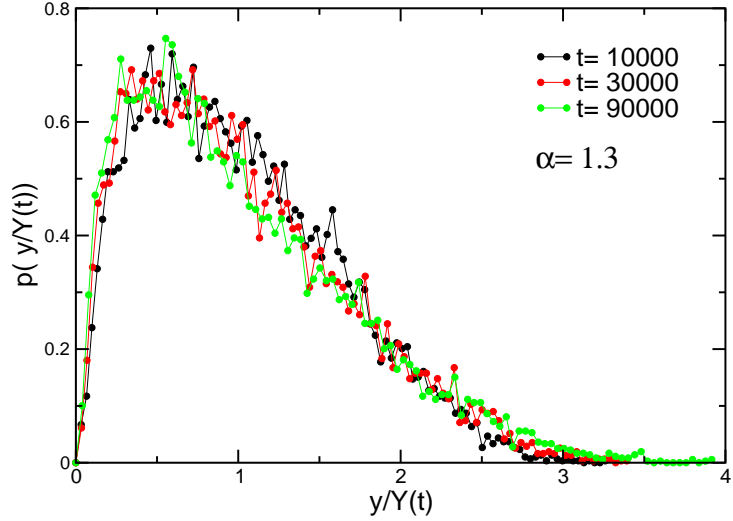}
\caption{The distribution of heights scales as $y/Y(t)$, where
$Y(t)$ is given by Eq. (\ref{Y-t}) for $\alpha= 1.3$. The
distribution is shown at three different times (black, red and green
curves), all within the limiting region of large times where
asymptotic behavior $y\propto t^{1-1/\alpha}$ holds.
}\label{fig:DistrYalpha13Scaled}
\end{figure}

Moreover, the integrated distribution $p(y,t)$ exhibits the
same scaling as $\rho(0,y,t)$. Indeed, using the relation
(\ref{F-y-hat-derivatives-relation}) in Eq.
(\ref{p-y-t-equation-1}) we obtain:
\begin{align}
p(y,t) &= p_0(y) + \frac{1}{Y(t)} \frac{\alpha}{\alpha - 1}
\frac{1}{\hat y} F(\hat y). \label{p-y-t-growing}
\end{align}
Fig. \ref{fig:DistrYalpha13Scaled} shows the scaling collapse
of the numerically calculated distributions $p(y,t)$ for
$\alpha = 1.3$ and three different times. We see that, indeed,
$p(y,t)$ is a scaling function of $y/Y(t)$ in agreement with
our predictions.

\begin{figure}
\centering
\includegraphics[scale=.4]{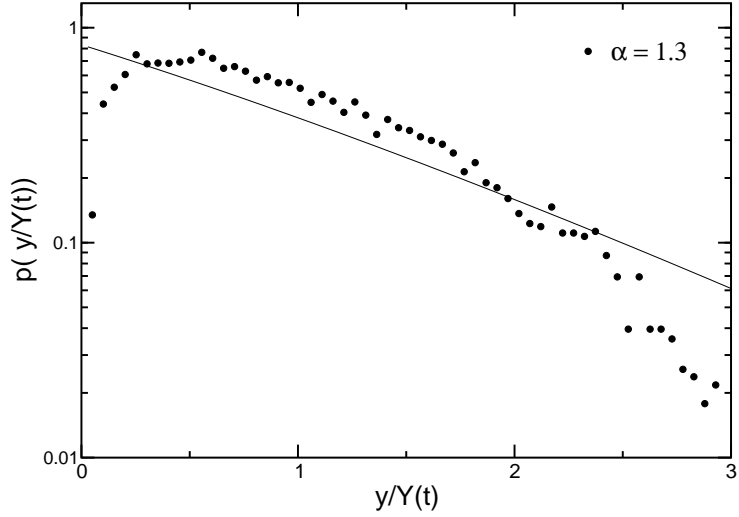}
\caption{The distribution $p(y,t)$ as a function of $y/Y(t)$.
The theoretical prediction Eq. (\ref{p-y-t-growing}) (solid
curve) and the numerical distribution (black dots) are
different, however, they have a similar dependence on $y$ for
the values where we believe the solution is valid, $t_0^{1/2}
\ll y \ll t^{1/2}$. Here, $Y(t)= 50$, $t^{1/2} \approx 300$ and
$t_0^{1/2} \approx 1$, so the region of validity of Eq.
(\ref{p-y-t-growing}) in the scaled variable is $0.02 \ll
y/Y(t) \ll 6$. This explains the disagreement between the
theory and the numerics for $y/Y(t) > 2$. Also, for small
values of $y/Y(t)$ where we should not trust Eq.
(\ref{p-y-t-growing}), the theory still gives a significant
weight to the distribution $p(y,t)$. This is, presumably, the
reason for the discrepancy between the numerics and the
theory in the range $y/Y(t) < 2$.}
\label{fig:DistrYalpha13CompareToFit}
\end{figure}

We can calculate the asymptotics of the function $F(\hat y)$.
For small values of $\hat y$ we can neglect the term with $\hat
y$ in the exponential in Eq. (\ref{F-y-hat}), as well as
replace the sine function under the integral by its (small)
argument:
\begin{align}
F(\hat y \ll 1) &\approx \frac{\alpha - 1}{\alpha}
\frac{1}{\Gamma\big(\tfrac{1}{\alpha}\big)} \hat y.
\label{F-y-hat-small}
\end{align}
For large $\hat y$ we need to use the steepest descent method
for the contour integral in Eq. (\ref{F-y-hat}), which results
in
\begin{align}
F(\hat y \gg 1) &\approx \Big(\frac{\alpha}{2\pi}\Big)^{1/2}
\Big(\frac{\alpha - 1}{\alpha} \hat y \Big)^{\alpha/2} \exp
\Big[-\frac{1}{\alpha-1}\Big(\frac{\alpha - 1}{\alpha} \hat y
\Big)^\alpha \Big]. \label{F-y-hat-large}
\end{align}
We have to remember that we can only trust this result for $y_0
\ll y \ll t^{1/2}$.

Figure \ref{fig:DistrYalpha13CompareToFit} shows a comparison
between the numerical data and the theoretical prediction of
Eq. (\ref{p-y-t-growing}) for the distribution $p(y,t)$. While
the overall dependence on $y$ is similar between the two, we
would obtain a better fit for $y_0 \ll y \ll t^{1/2}$ if we
redistributed the weight outside this region to the range were
Eq. (\ref{p-y-t-growing}) is valid.


\begin{figure}
\centering
\includegraphics[scale=.4]{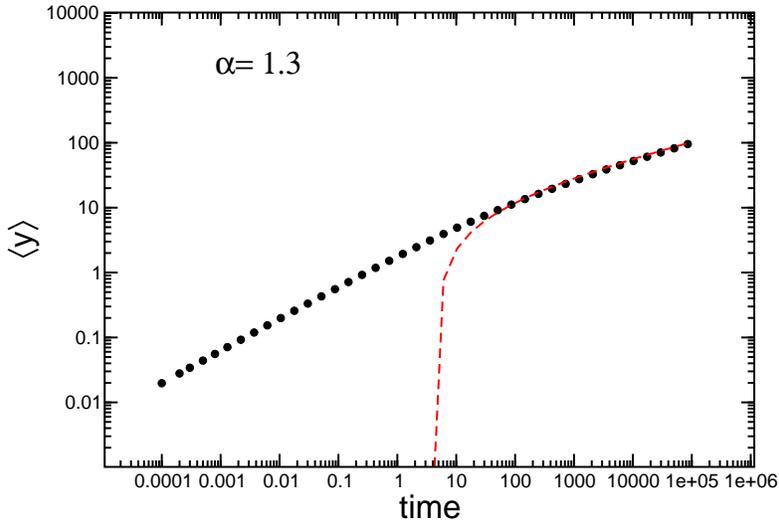}
\caption{The average height $\langle y\rangle$ for SLE driven
by L\'evy flights with $\alpha= 1.3$ grows as a power-law
$t^{1-1/\alpha}$. The red dashed line is a fit to Eq.
(\ref{y-average-alpha>1}) for $t>1$, where we only vary the
parameter $y_0$.} \label{fig:AvgYalpha13}
\end{figure}

Next, we will calculate the time evolution of the average
height of the growing trees $\langle y \rangle$ from Eqs.
(\ref{y-average}, \ref{normalization-alpha>1}):
\begin{align}
\langle y \rangle &= y_0 + \frac{1}{\Gamma(1 - \tfrac{1}{\alpha})}
\int_{0}^t \!\! dt' \frac{Y(t')}{t'} =
y_0 + \frac{2}{c^{1/\alpha}} \frac{\Gamma(1 +
\tfrac{1}{\alpha})}{1 - \tfrac{1}{\alpha}}
t^{1-\frac{1}{\alpha}}. \label{y-average-alpha>1}
\end{align}
Here, all short time contributions are included in $y_0$. This
nicely fits the numerics, see Fig. \ref{fig:AvgYalpha13}, and
reproduces the result (\ref{hand_waving2}) of the simple
argument using the Langevin equation.

\begin{figure}
\centering
\includegraphics[scale=.4]{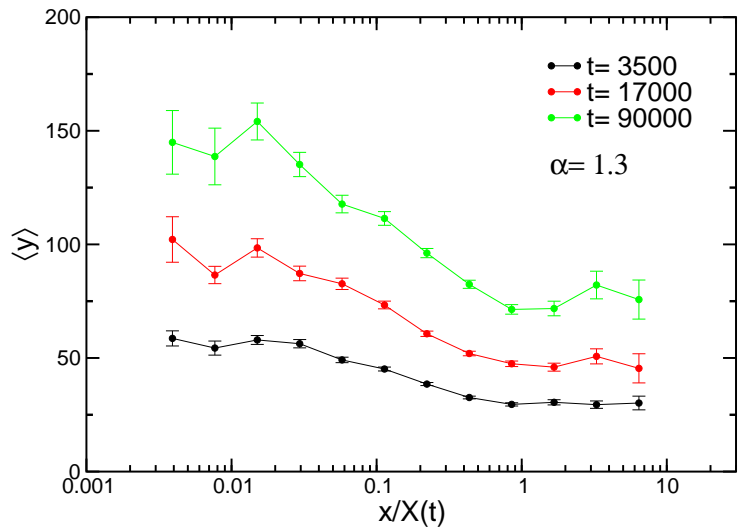}
\caption{Average height of the trace $y=\mathrm{Im}\,\gamma(t)$
as a function of $x/(ct)^{1/\alpha}$ for SLE driven by Levy
flights with $\alpha= 1.3$. $y$ data are bined logarithmically.
Here, $(ct)^{1/\alpha}=\ 1827.15,\ 6217.82,\ 21156.6$ for the
three values of time. The average height close to $x=0$ is $2$
times bigger than the height at large $x$ and roughly $1.4$
times higher than the global average $\langle y \rangle$. The
theoretically predicted value of the ratio between the height
at $x=0$ and the average is $1.9$ (Eq. (\ref{y0-y-ratio})).
This discrepancy is most probably due to a finite time effect
and the limited amount of data close to
$x=0$.}\label{fig:AvgYvsXalpha13}
\end{figure}

We also want to compare the distribution at $x=0$ to the
distribution averaged over all $x$. We calculate the average
value $\langle y \rangle^0$ (the superscript indicates that
this average is calculated at $x=0$) from the distribution
(\ref{rho-0-y-t-alpha>1-normalized}):
\begin{align}
\langle y \rangle^0 &= \frac{4}{\alpha c^{1/\alpha}} \Big|\cos
\frac{\pi}{\alpha}\Big| \Gamma\Big(1 - \frac{1}{\alpha}\Big)
\Gamma\Big(\frac{2}{\alpha} -1 \Big) t^{1 - 1/\alpha}.
\label{y-average-alpha>1-x=0}
\end{align}
The ratio of the two averages (neglecting $y_0$) is
\begin{align}
\frac{\langle y \rangle^0}{\langle y \rangle} = \frac{1}{\pi}
\Big|\sin \frac{2\pi}{\alpha} \Big| \Gamma\Big(1 -
\frac{1}{\alpha}\Big) \Gamma\Big(\frac{2}{\alpha} - 1\Big)
\Gamma\Big(2 - \frac{1}{\alpha}\Big). \label{y0-y-ratio}
\end{align}
This tends to 2 as $\alpha \to 1$ from above, and to $\pi/2$ as
$\alpha \to 2$ from below. We observe similar behavior in our
numerical results, where the average of $y$ at $x=0$  is higher
than the overall average (Fig. \ref{fig:AvgYvsXalpha13}).
However, the ratio (\ref{y0-y-ratio}) is not matched exactly.
Presumably, this is because we do not have enough data close to
$x=0$ and we cannot reach long enough times in order for the
various constants (like $y_0$) to be negligible, so that Eq.
(\ref{y0-y-ratio}) is accurate.

\subsection{Results for $\alpha = 1$}

Now we use the approximations (\ref{K-lambda-small-alpha=1},
\ref{K-lambda-y-alpha=1}) from Appendix
\ref{app:XK-asymptotics} in Eq.
(\ref{solution-0-y-lambda-simplified}). The resulting
expression for $\rho(0,y,\lambda)$ is difficult to analyze
without further approximations. We will evaluate it as well as
its inverse Laplace transform with logarithmic accuracy, which
amounts to three assumptions. First, we assume that all the
logarithms that appear are large compared to constants of order
one such as $\pi$, $c$, etc, which will be neglected. Secondly,
the logarithms are assumed to be small compared to power laws
for large arguments: $\ln t \ll t$. Finally, the logarithms are
slow functions as compared to power laws and exponentials, and
in integrals can be replaced by their values at the typical
scale of variation of the fastest function under the integral.
All subsequent equations in this section will be obtained with
logarithmic accuracy using these assumptions.

First we have
\begin{align}
\rho(0,y, \lambda) \approx \frac{1}{2 \pi} \frac{\ln
\frac{1}{\lambda t_0} \ln \frac{c}{\lambda y_0}} {\ln^2
\frac{c}{\lambda y}} \exp \bigg( -\frac{c}{2} \frac{y}{\ln
\frac{c}{\lambda y}} \bigg). \label{rho-0-y-lambda-alpha=1}
\end{align}
The time dependence now follows from the inverse Laplace
transform, using the same contour integral described in the
previous section:
\begin{align}
\rho(0,y,t) \approx \frac{1}{2 \pi^2 t} \int_0^\infty \!\!
d\lambda \frac{\ln\frac{t}{\lambda t_0} \ln\frac{c t}{\lambda
y_0}} {\ln^2 \frac{c t}{\lambda y}} \sin\bigg(\frac{\pi}{2}
\frac{cy}{\ln^2 \frac{c t}{\lambda y}} \bigg) \exp \bigg( \!\!
-\lambda -\frac{c}{2} \frac{y}{\ln \frac{c t}{\lambda y}}
\bigg). \label{rho-0-y-t-alpha=1}
\end{align}
The integral of this expression over $y$
\begin{align}
\int_0^\infty \!\! dy \, \rho(0,y,t) & \approx \frac{1}{\pi c
t} \frac{\ln \frac{t}{t_0} \ln \frac{c t}{y_0}} {\ln^2
\big(\frac{c^2 t}{2}\big)} \label{normalization-alpha=1}
\end{align}
leads to a normalized distribution at $x=0$:
\begin{align}
\rho_n(0,y,t) &\approx \frac{c}{2 \pi}
\frac{\ln^2 \big(\frac{c^2 t}{2}\big)}
{\ln \frac{t}{t_0} \ln \frac{c t}{y_0}}
\nonumber \\
&\quad \times \int_0^\infty \!\! d\lambda
\frac{\ln\frac{t}{\lambda t_0} \ln\frac{c t}{\lambda y_0}}
{\ln^2\frac{c t}{\lambda y}} \sin\bigg(\frac{\pi}{2}
\frac{cy}{\ln^2 \frac{c t}{\lambda y}} \bigg) \exp \bigg( \!\!
-\lambda -\frac{c}{2} \frac{y}{\ln \frac{c t}{\lambda y}}
\bigg). \label{rho-0-y-t-alpha=1-normalized}
\end{align}

\begin{figure}
\centering
\includegraphics[scale=.4]{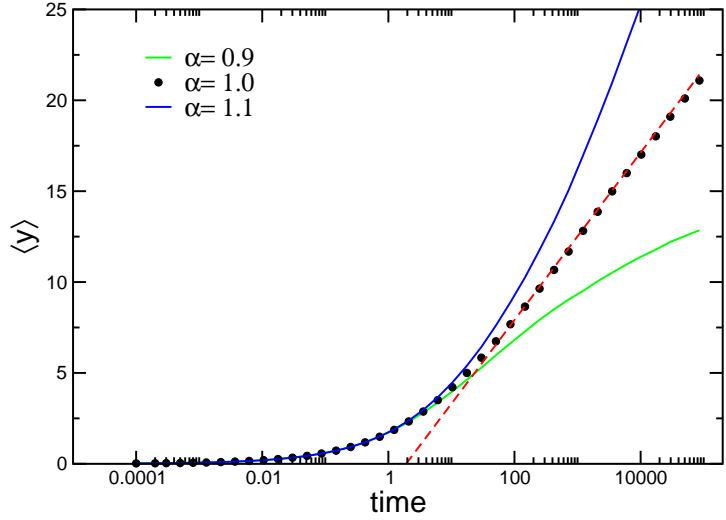}
\caption{The average height $\langle y\rangle$ for SLE driven
by L\'evy flights $\alpha= 1.0$ grows logarithmically with
time. The red dashed line is a one parameter fit for $t>10$ to
the predicted function $A+\frac{2}{c}\ln{t}$.}
\label{fig:AvgYalpha1}
\end{figure}

The mean value of the height of trees near $x=0$ follows from
$\rho_n(0,y,t)$ using the same arguments as before:
\begin{align}
\langle y \rangle^0 &\approx
\frac{4}{c} \ln\frac{c^2 t}{2}.
\label{y-average-alpha=1-x=0}
\end{align}
The average height (over all $x$) is also found easily from Eq.
(\ref{y-average}):
\begin{align}
\langle y \rangle &= y_0 + \frac{2}{c} \int_{t_0}^t \frac{dt'}{t'}
\frac{\ln\frac{t'}{t_0} \ln \frac{c t'}{y_0}}
{\ln^2 \big(\frac{c^2 t'}{2}\big)}
\approx \frac{2}{c} \ln t + \text{ const}.
\label{y-average-alpha=1}
\end{align}
As shown in Fig. \ref{fig:AvgYalpha1} this is in good agreement
with the numerics. The ratio of the two averages in the long
time limit is $\langle y \rangle^0/\langle y \rangle = 2$,
consistent with the limit $\alpha \to 1$ of Eq.
(\ref{y0-y-ratio}).

The asymptotics of the distribution $\rho_n(0,y,t)$ for small
and large values of $y/\ln ct$ can be found similar to the case
$\alpha > 1$:
\begin{align}
\rho_n(0,y,t) &\approx
\begin{cases}
\dfrac{c^2}{4} \dfrac{\ln^2 \frac{c^2 t}{2}}{\ln^4 \frac{c
t}{y}} y, & y_0 \ll y \ll \ln ct, \\
\dfrac{c^{3/2}}{4 \pi^{1/2}} \dfrac{\ln^2 \frac{c^2 t}{2}} {\ln
\frac{t}{t_0} \ln \frac{c t}{y_0}} \dfrac{\ln \frac{8t}{c t_0
y} \ln \frac{8t}{y_0 y}} {\ln^3 \frac{8t}{y^2}} y^{1/2} \exp
\Big(-\dfrac{cy}{2\ln \frac{8t}{y^2}}\Big), & \ln ct \ll y \ll
t^{1/2}.
\end{cases}
\label{rho-0-y-small-t-alpha=1-normalized}
\end{align}

Finally, using Eq. (\ref{p-y-t-equation-1}), we get an
expression for the integrated distribution:
\begin{align}
p(y,t) &\approx p_0(y) + \frac{c}{2\pi} \ln\frac{t}{t_0}
\int_0^\infty \!\! d\lambda \frac{\ln\frac{t}{\lambda t_0}
\ln\frac{c t}{\lambda y_0}} {\ln^3\frac{c t}{\lambda y}}
\sin\bigg(\frac{\pi}{2} \frac{cy}{\ln^2\frac{c t}{\lambda y}}
\bigg) \exp \bigg( \!\! -\lambda -\frac{c}{2} \frac{y}{\ln
\frac{c t}{\lambda y}} \bigg). \label{}
\end{align}
The asymptotics of these expression follow as before:
\begin{align}
p(y,t) - p_0(y) &\approx
\begin{cases}
\dfrac{c^2}{4} \dfrac{\ln^2\frac{t}{t_0} \ln\frac{c t}{y_0}}
{\ln^5 \frac{c t}{y}} y, & y_0 \ll y \ll \ln ct, \\
\dfrac{c^{3/2}}{4 \pi^{1/2}} \dfrac{\ln\frac{t}{t_0}
\ln\frac{8t}{c t_0 y} \ln\frac{8t}{y_0 y}}
{\ln^4\frac{8t}{y^2}} y^{1/2} \exp
\Big(-\dfrac{cy}{2\ln\frac{8t}{y^2}}\Big), & \ln ct \ll y \ll
t^{1/2}.
\end{cases}
\label{}
\end{align}

\subsection{Results for $\alpha < 1$}

\begin{figure}
\centering
\includegraphics[scale=.4]{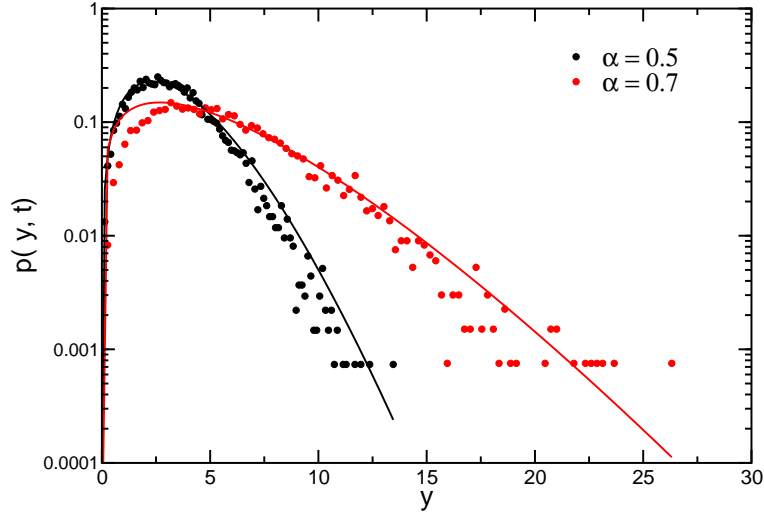}
\caption{Distribution fits for $\alpha<1$. We compare the
numerically calculated distribution $p(y,t)$ to the theoretical
curve for $\rho_\infty$ given by Eq.
(\ref{rho-0-y-t-alpha<1-stationary}) with one free parameter
for normalization. We claim that $p(y,t)=\rho_\infty(0,y)$ for
$y_0 \ll y \ll t^{1/2}$ where our solution is
valid.}\label{fig:DistrYsaturatedCompareToFits}
\end{figure}

In this case we use the approximation (\ref{K-approx-a<1}) from
Appendix \ref{app:XK-asymptotics} leading to
\begin{align}
\rho(0, y, \lambda) &\approx  N \, K(\lambda) y^{2 - 2\alpha}
\exp \Big( -\frac{y^{2-\alpha}}{C(2-\alpha)} \Big),
\label{rho-0-y-lambda-alpha<1-0} \\
C &= \frac{2}{c} \Gamma(1-\alpha), \qquad N=\frac{y_0^{\alpha -
1}}{2\pi C}.
\end{align}
The inverse Laplace transform of this expression gives the
leading approximation
\begin{align}
\rho(0, y, t) \approx \frac{N}{X(t)} y^{2 - 2\alpha} \exp \Big(
-\frac{(1 - \alpha)c}{2\Gamma(3-\alpha)} y^{2-\alpha}\Big).
\label{rho-0-y-t-alpha<1-0}
\end{align}

The obtained result depends on time only through the overall
factor $X^{-1}(t)$. We can understand this as follows. The
distribution $\rho(x, y, t)$ at long times spreads in the $x$
direction up to the scale $X(t)$ but becomes stationary in the
$y$ direction. Therefore, at the particular value $x=0$ the
density $\rho(0, y, t)$ decays with time as $X^{-1}(t)$.
However, if we are looking at the distribution of the $y$
coordinate for $x=0$, we should normalize Eq.
(\ref{rho-0-y-t-alpha<1-0}) which gives the truly stationary
distribution (normalized by the appropriate choice of $N_1$)
\begin{align}
\rho_\infty(0, y) \approx N_1 y^{2 - 2\alpha} \exp \Big(
-\frac{(1 - \alpha)c}{2\Gamma(3-\alpha)} y^{2-\alpha}\Big),
\label{rho-0-y-t-alpha<1-stationary}
\end{align}
in agreement with numerics, see Fig.
\ref{fig:DistrYsaturatedCompareToFits}, where we actually
observe that the integrated distribution $p(y,t)$ coincides
with $\rho_\infty(0, y)$ at long times.

\begin{figure}
\centering
\includegraphics[scale=0.32]
{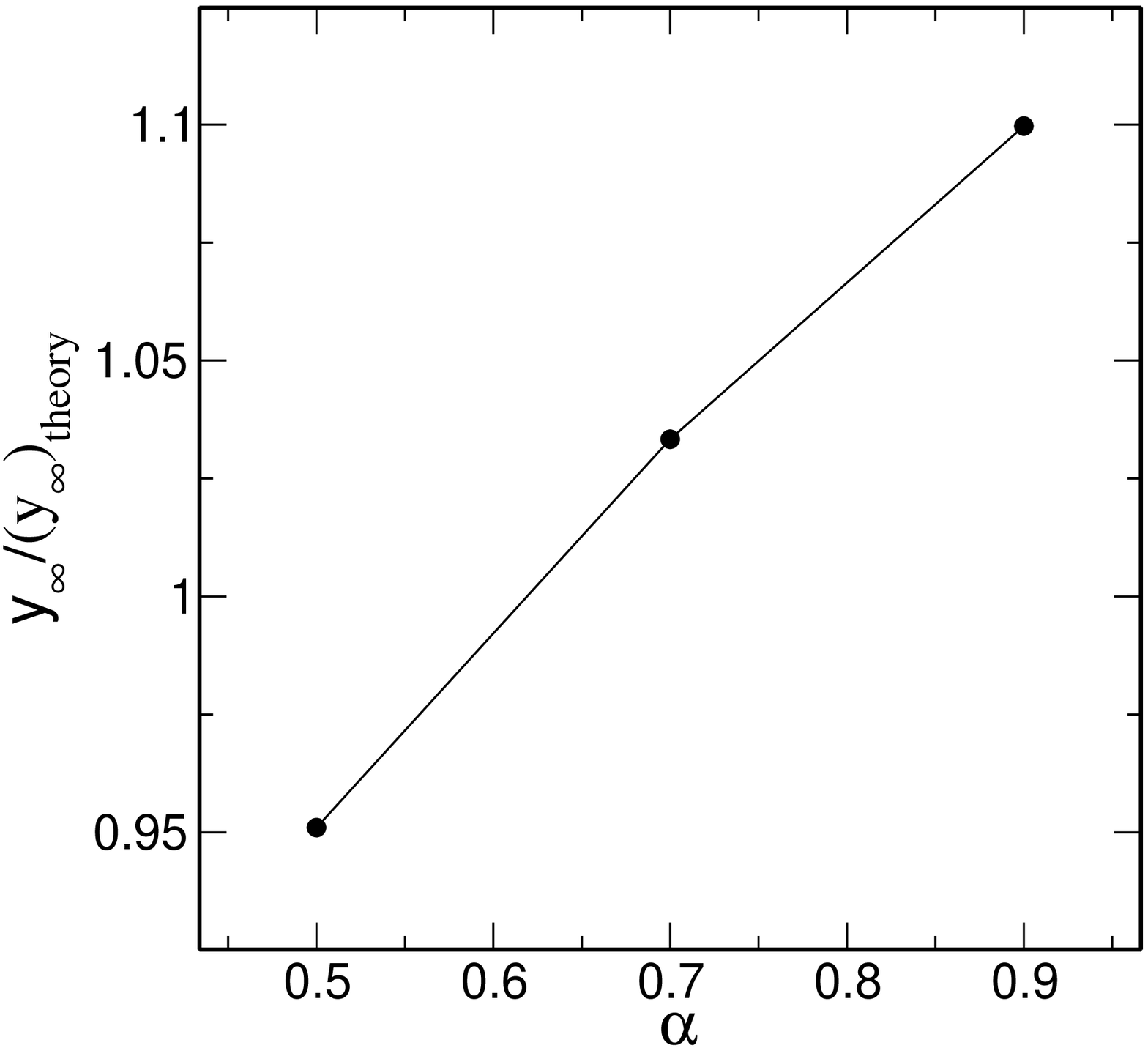} \hfill
\includegraphics[scale=0.32]{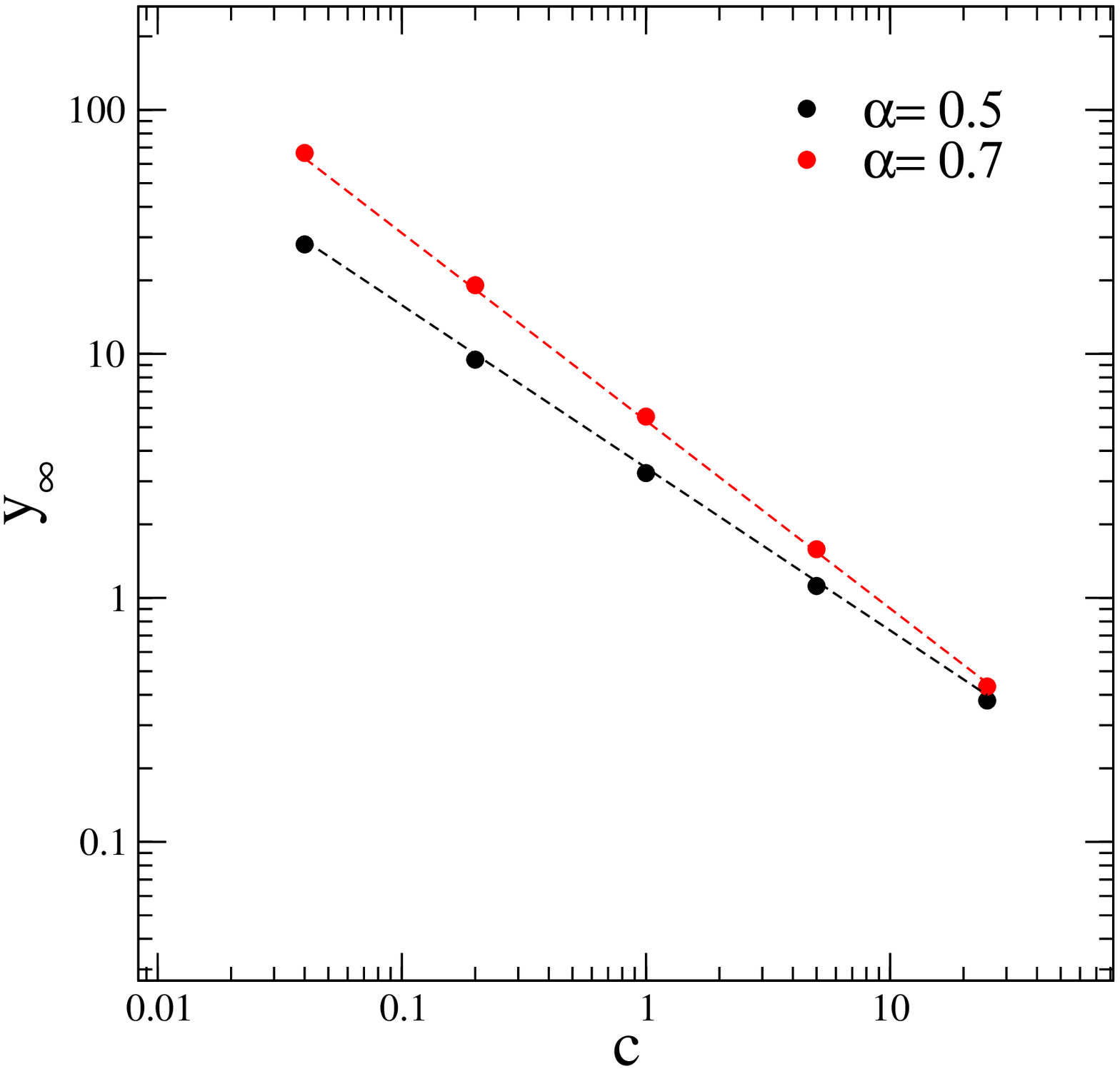}
\caption{(Left) The ratio of the numerically calculated over
the theoretically predicted value of the saturated height
$y_\infty$ for $\alpha<1$. (Right) The saturated height of the
trees vs. the strength. The dashed lines are the theoretical
values for $y_\infty$, Eq. (\ref{y-infinity}).}
\label{fig:FinalYsaturated}
\end{figure}

The stationary distribution
(\ref{rho-0-y-t-alpha<1-stationary}) allows us to calculate the
average saturated height of the trees:
\begin{align}
y_\infty = \int_0^\infty dy \, y \rho_\infty(0,y) = \Big(\frac{2
\Gamma(3-\alpha)}{1- \alpha} \Big)^{1/(2-\alpha)}
\Gamma^{-1}\Big(\frac{3-2\alpha}{2-\alpha}\Big) c^{-1/(2-\alpha)}.
\label{y-infinity}
\end{align}
This is in very good agreement with the numerically calculated
values shown in Fig. \ref{fig:FinalYsaturated}.

\begin{figure}
\centering
\includegraphics[scale=.4]{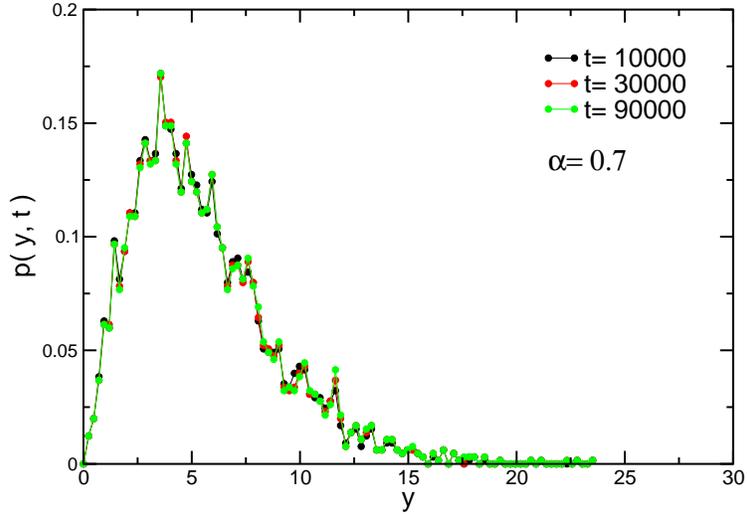}
\caption{Distribution of heights averaged over all $x$ for SLE
driven only by L\'evy flights and $\alpha= 0.7$. The distribution is
shown at three different times (black, red and green curves),
corresponding at the limit of large time. We see how the
distribution of the height of trees is stationary.
}\label{fig:DistrYalpha07}
\end{figure}

Let us discuss now the integrated distribution $p(y,t)$ and its
mean $\langle y \rangle$. Unfortunately, in the present case
($\alpha < 1$), the Eqs. (\ref{p-y-t-equation-1}) and
(\ref{y-average}) do not give reliable results simply because
the apparent distribution and saturation height are very
sensitive to the lower limit $t_0$, and the results are of the
same order as the initial conditions at $t_0$. Analytically, we
can see that the distribution $p(y,t)$ becomes stationary as $t
\to \infty$, even though we cannot determine $p(y,\infty)$. The
time independence of the distribution $p(y,t)$ at long times is
checked numerically in Fig. \ref{fig:DistrYalpha07}. Numerics
presented in Fig. \ref{fig:DistrYsaturatedCompareToFits}
indicate that $p(y,t)=\rho_\infty(0,y)$ (see Eq.
(\ref{rho-0-y-t-alpha<1-stationary})) for the appropriate range
$y_0 \ll y \ll t^{1/2}$, and we will discuss why this is true
below.

\begin{figure}
\centering
\includegraphics[scale=.4]{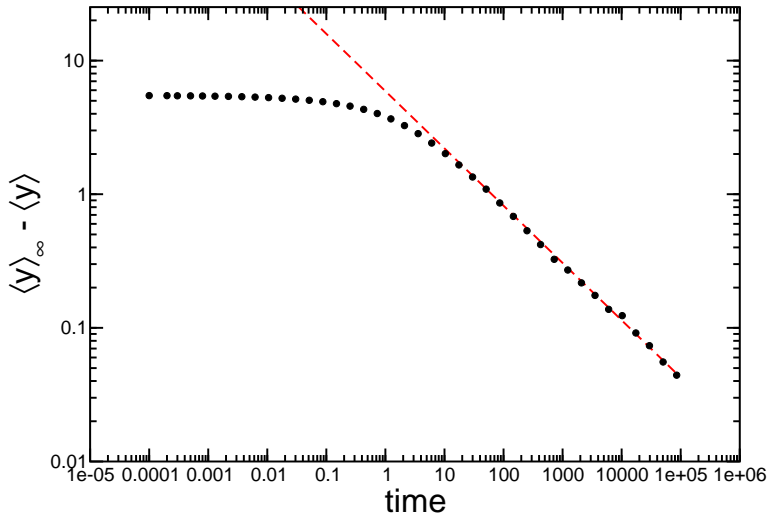}
\caption{The average height $\langle y\rangle$ for SLE driven
by L\'evy flights $\alpha= 0.7$ saturates to
$\langle y\rangle_\infty$ as $t^{1-1/\alpha}$. The red dashed
line is the analytic result, Eq. (\ref{y-infinity-approach}),
with the value of $D =
\frac{2\Gamma(1+1/\alpha)}{c^{1/\alpha}(1-1/\alpha)}$ obtained
in
Eq. (\ref{hand_waving2}). $\langle y\rangle_\infty$ was
calculated from the two numerical points of $y$ for the largest
times.} \label{fig:AvgYalpha07}
\end{figure}

We can also see that the way the average tree height approaches
its limiting value is given by the power law
\begin{align}
\langle y \rangle &= \langle y \rangle_\infty - D c^{-1/\alpha}
t^{1 - 1/\alpha}. \label{y-infinity-approach}
\end{align}
We have previously calculated $D$ in Eq. (\ref{hand_waving2})
using the Langevin formulation of the process. This result
agrees well with numerics, as demonstrated in Fig.
\ref{fig:AvgYalpha07}.

\begin{figure}
\centering
\includegraphics[scale=.4]{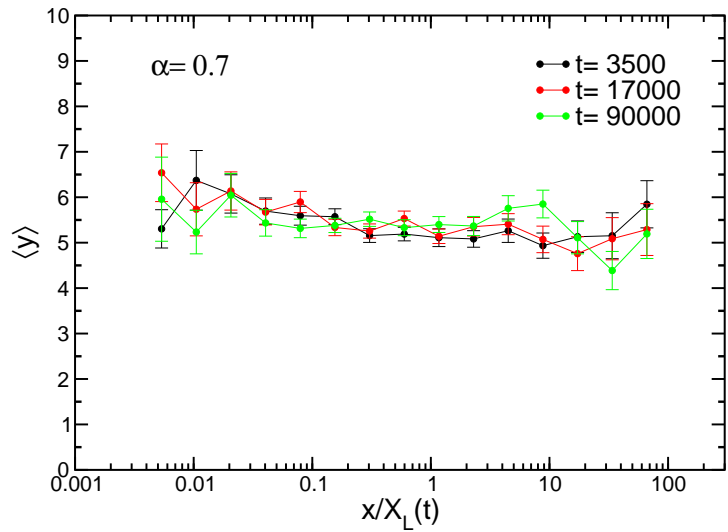}
\caption{Average height of the trace $y=\mathrm{Im}\,\gamma(t)$ as a
function of $x/(ct)^{1/\alpha}$ for SLE driven by Levy flights with
$\alpha= 0.7$. $y$ data are binned logarithmically and the average of
every bin is plotted. $(ct)^{1/\alpha}=\ 2.9\ 10^5, 2.8\ 10^6, 2.75\
10^7$.}
\label{fig:AvgYvsXalpha07}
\end{figure}

We can argue that $\langle y \rangle_\infty = y_\infty$ and
that $p(y,\infty) = \rho_\infty$, if we return to the initial
description of the process seen as SLE trees growing forward in
time \cite{previous paper}. For $\alpha<1$ the jumps of the
L\'evy process are large and we know that any new tree is most
likely to grow starting from the real axis. The trees are
sparse and the new tree will grow isolated from its neighbors,
hence it will be identical to any other tree, including the
trees that grow close to the origin at $x=0$. Therefore, we
expect the distribution of $y$'s at $x=0$ to be identical to
the distribution at any other $x$. Numerics also support this
argument. In Fig. \ref{fig:AvgYvsXalpha07} we observe that the
average height of the trees is practically independent of the
value of $x$. Also, in Fig.
\ref{fig:DistrYsaturatedCompareToFits} we compare $p(y,\infty)$
and $\rho_\infty$ while in Fig. \ref{fig:FinalYsaturated} we
show that $\langle y \rangle_\infty = y_\infty$.

\section{Conclusions}
\label{sec:conclusions}

In this paper we have analyzed the global properties of growth
in the complex plain described by a generalized stochastic
Loewner evolution driven by a symmetric stable L\'evy process
$L_\alpha(t)$, introduced in our previous paper \cite{previous
paper}. The phase transition at $\alpha = 1$ whose implications
for local properties of growth were the subject of Ref.
\cite{previous paper}, also manifests itself on the whole plane
resulting in a rich scaling behavior.

We have used a Fokker-Planck equation to study the joint
distribution $\rho(x,y,t)$ for the real and imaginary parts of
the tip of the growing trace. The presence of the L\'evy
flights in the driving force imposes very different dynamics in
the $x$ and $y$ directions. While in the $x$ direction the
process spreads similarly to the L\'evy forcing $x \sim X(t)
\sim t^{1/\alpha}$, the SLE dictates $y \ll X(t)$, for all
values of $\alpha$. This separation of the horizontal and
vertical scales in the process allows us to make sensible
approximations and explore geometric properties of the
stochastic growth in all phases, $\alpha < 1$, $\alpha = 1$,
and $\alpha > 1$, both qualitatively and quantitatively.

For $\alpha<1$, the vertical growth saturates at a finite
height $y_\infty$. In terms of the picture presented in
\cite{previous paper}, long jumps occur often so that new trees
grow isolated and there is a small chance that the trace grows
on an already existing tree.

For $\alpha > 1$, the average height of the process grows as a
power law $t^{1 - 1/\alpha}$ with time. New trees grow close to
old ones, so that when the process returns to a previously
visited part of the real axis it will have to grow on top of
already existing trees. Eventually the trace will grow past any
point on the plane.

At the boundary between the two phases, $\alpha=1$, the height
of the process grows logarithmically with time.

\section{Acknowledgements}

This research was supported in part by NSF MRSEC Program under
DMR-0213745. IG was also supported by an award from Research
Corporation and the NSF Career Award under DMR-0448820. We wish
to acknowledge many helpful discussions with Paul Wiegmann,
Eldad Bettelheim, and Seung Yeop Lee. IG also acknowledges
useful communications with Steffen Rohde.

\appendix

\setcounter{section}{0}

\section{Appendices}

\subsection{Numerical calculations}
\label{app:numerics}

The interpretation of equation (\ref{Langevin-phys}) is very
helpful to our calculations. $z_t$ and the tip of the trace
have the same distribution. This allows, instead of calculating
the trace $\gamma(t)$ for every time $t$ and noise realization
($O(n^2)$), to efficiently collect statistics for the position
of the tip by integrating the Langevin equation
(\ref{Langevin-phys}) ($O(n)$).

Following Ref. \cite{previous paper} we approximate $\xi(t)$ by
a piecewise constant function with jumps appropriately
distributed: $\xi(t)= \xi_{j}$ for $(j - 1)\tau<t<j\tau$. For
such a driving function the process $z_t$ in Eq.
(\ref{Langevin-phys}) can then be calculated numerically as an
iteration process of infinitesimal maps \cite{hastings}
starting from the condition $z = 0$ as follows:
\begin{align}
\label{eq:ITERATE-MAP} z_{n} = z(n\tau)=f_{n}\circ\ f_{n -
1}\ ...\circ\ f_{1} (0) - \xi_n.
\end{align}
The infinitesimal conformal map $f_n$ at each time interval $n$
is defined by:
\begin{align} \label{eq:INV-MAP}
f_{n}(z)=w_n^{-1}(z) = \sqrt{(z - \xi_n)^2 - 4\tau} + \xi_n
\end{align}
The value of $\xi_n$ is randomly drawn from the appropriate
distribution. The number of steps necessary to produce an SLE
trace up to step $n$ grows only as $O(n)$. All numerical
results in the next section have been calculated using the
average of Eq. (\ref{eq:INV-MAP}) over many noise realizations.

The trace can also be produced directly \cite{previous paper},
as $g^{-1}(\xi(t),t)$, in which case we approximate
\begin{align}
\gamma_j = \gamma(j\tau) = f_1\ldots\circ f_{n-1}\circ f_n(\xi_n).
\label{eq:ITERATE-TRACE}
\end{align}
However, the number of steps in this method grows as $O(n^2)$.
We used this method to verify that numerically calculated $z$
and $\gamma$ have identical distributions. Eq.
(\ref{eq:ITERATE-TRACE}) was also used to calculate the traces
shown in Fig. \ref{fig:trees}.

Here, we will assume $\kappa=0$ for simplicity, that is, the
driving force is pure L\'evy flights
$\xi(t)=c^{1/\alpha}L_\alpha(t)$. The addition of a Brownian
motion will not affect our conclusions. For all realizations of
the L\'evy-SLE process we take $c=1$ and $\tau = 10^{-4}$
unless otherwise noted.

\subsection{Asymptotics for $K(\lambda, y)$}
\label{app:XK-asymptotics}

Let us consider (we need to use the lower cut off $t_0$ here to
have a convergent result for $\alpha < 1$)
\begin{align}
K(\lambda) &= \int_{t_0}^\infty \!\! dt \, \frac{e^{-\lambda
t}}{X(t)} = \frac{2\Gamma\big(1 +
\frac{1}{\alpha}\big)}{c^{1/\alpha}} \int_{t_0}^\infty \!\! dt \,
t^{-1/\alpha} e^{-\lambda t}
\nonumber \\
&= \begin{cases} \dfrac{2\Gamma\big(1 +
\frac{1}{\alpha}\big)}{c^{1/\alpha}} \lambda^{-1 +1/\alpha}
\Gamma\big(1 - \tfrac{1}{\alpha}, \lambda
t_0 \big), & \alpha \neq 1, \\
\dfrac{2}{c} E_1(\lambda t_0), & \alpha = 1,
\end{cases}
\label{K-lambda}
\end{align}
where $\Gamma(a,x)$ is the incomplete gamma function, and
$E_1(x)$ is the exponential integral. Since $\lambda$ has the
dimension and the meaning of frequency, and we are interested
in $t \gg t_0$, we will only need the small argument
asymptotics of these functions:
\begin{align}
\Gamma(a,x) &\approx \Gamma(a) - \frac{x^a}{a}, &
E_1(x) &\approx  - \ln x , & x &\ll 1, \label{}
\end{align}
This gives for $\lambda t_0 \ll 1$
\begin{align}
K(\lambda) &\approx A \lambda^{-1 +1/\alpha} + B t_0^{1 - 1/\alpha},
& \alpha \neq 1,
\label{K-lambda-small-general-alpha}\\
& A = \frac{2\pi}{\alpha c^{1/\alpha} \sin\frac{\pi}{\alpha}},
&B = \frac{2}{c^{1/\alpha}} \frac{\alpha}{1- \alpha}
\Gamma\big(1 + \tfrac{1}{\alpha}\big), \\
K(\lambda) &\approx \dfrac{2}{c} \ln\frac{1}{\lambda t_0}, & \alpha
= 1, \label{K-lambda-small-alpha=1}
\end{align}
For $\alpha > 1$ we can set $t_0 = 0$ and obtain
\begin{align}
K(\lambda) &= A \lambda^{-1 +1/\alpha}, & \alpha > 1
\label{K-lambda-alpha>1}
\end{align}
and for $\alpha < 1$ we can set $\lambda = 0$:
\begin{align}
K(0) &= B t_0^{1 - 1/\alpha}, & \alpha < 1. \label{K-0-alpha<1}
\end{align}

We now turn to the Laplace transform $K(\lambda, y)$:
\begin{align}
K(\lambda, y) &= 2 \int_0^\infty \!\! dk \, \frac{e^{-yk}}{\lambda +
c k^\alpha}. \label{K_0-lambda-y}
\end{align}
Since in the Laplace transform the important values of
$\lambda$ are the inverse typical time scales, this means that
the relevant asympotics of $K(\lambda, y)$ are those with
$\lambda y^\alpha/c \ll 1$. The opposite case of $\lambda
y^\alpha/c \gg 1$ corresponds to short times, where our basic
approximation is invalid. So from now on we will focus on the
limit $\lambda y^\alpha/c \ll 1$.

This integral can be evaluated exactly in a number of cases.
First, when $y = 0$, the integral converges for $\alpha > 1$
and gives the same expression as $K(\lambda)$ in Eq.
(\ref{K-lambda-alpha>1}). Secondly, for $\lambda = 0$ the
integral converges (for $y > 0$) for $\alpha < 1$ and gives
then
\begin{align}
K(0, y) &= \frac{2}{c} \int_0^\infty \!\! dk \, k^{-\alpha} e^{-yk}
= C y^{\alpha - 1}, & C = \frac{2}{c} \Gamma(1-\alpha).
\label{K-0-y}
\end{align}
All the constants $A$, $B$, and $C$ defined above diverge as
$1/(\alpha - 1)$ as $\alpha \to 1$. Finally, for $\alpha = 1$
we get
\begin{align}
K(\lambda, y) &= \frac{2}{c}e^{\lambda y/c} E_1\big(\tfrac{\lambda
y}{c} \big) \approx
\dfrac{2}{c} \ln \dfrac{c}{\lambda y},
&  \dfrac{\lambda y}{c} \ll 1.
\label{K-lambda-y-alpha=1}
\end{align}

In general for $\lambda y^\alpha/c \ll 1$, a good approximation
for $K(\lambda, y)$ is the sum of expressions in Eqs.
(\ref{K-lambda-alpha>1}, \ref{K-0-y}):
\begin{align}
K(\lambda, y) &\approx A \lambda^{-1 + 1/\alpha} + C  y^{\alpha -
1}. \label{K-lambda-y-approx}
\end{align}
Not only this approximation reproduces the correct limits in
Eqs. (\ref{K-lambda-alpha>1}) and (\ref{K-0-y}), but in the
limit $\alpha \to 1$ it also reduces to Eq.
(\ref{K-lambda-y-alpha=1}). This approximation can be obtained
by splitting the $k$ interval in the integral in Eq.
(\ref{K_0-lambda-y}) into two at the value $k_0 =
(\lambda/c)^{1/\alpha}$ and in each resulting integral replace
the denominator by the largest term in it.

Notice that for $\alpha > 1$, and in the limit of interest
$\lambda y^\alpha/c \ll 1$ the first term in Eq.
(\ref{K-lambda-y-approx}) is much greater than the second, and
we can use Eq. (\ref{K-lambda-alpha>1}) for both $K(\lambda)$
and $K(\lambda, y)$:
\begin{align}
K(\lambda) &\approx K(\lambda, y) \approx A \lambda^{-1 +1/\alpha},
& \alpha > 1.
\label{K-approx-a>1}
\end{align}
For $\alpha < 1$ the opposite is true, and we can use Eq.
(\ref{K-0-y}) as a valid approximation:
\begin{align}
K(\lambda, y) &\approx C y^{\alpha - 1}, & C = \frac{2}{c} \Gamma(1-\alpha).
\label{K-approx-a<1}
\end{align}

\end{document}